\documentstyle{article}

\def\noheaderplainsetup{%

      \topmargin=0pt \headheight=0pt \headsep=0pt  
      \oddsidemargin=0pt \evensidemargin=0pt       
      \textheight=8.9truein \textwidth=6.5truein}  

\noheaderplainsetup

\begin{document}


\newcommand{\chess}{\mbox{\em Chess}}
\newcommand{\checkers}{\mbox{\em Checkers}}
\newcommand{\rneg}{\neg}               
\newcommand{\pneg}{\neg}               
\newcommand{\emptyrun}{\langle\rangle} 
\newcommand{\oo}{\bot}            
\newcommand{\pp}{\top}            
\newcommand{\xx}{\wp}               
\newcommand{\leg}[3]{\mbox{\bf Lm}{#2}^{#1}_{#3}} 
\newcommand{\legal}[2]{\mbox{\bf Lr}^{#1}_{#2}} 
\newcommand{\Legal}[1]{\mbox{\bf LR}^{#1}} 
\newcommand{\win}[2]{\mbox{\bf Wn}^{#1}_{#2}} 
\newcommand{\tree}[2]{\mbox{\em Tree}^{#1}_{#2}} 
\newcommand{\constants}{\mbox{\bf Constants}} 
\newcommand{\variables}{\mbox{\bf Variables}}    
\newcommand{\moves}{\mbox{\bf Moves}}        
\newcommand{\runs}{\mbox{\bf Runs}} 
\newcommand{\players}{\mbox{\bf Players}} 
\newcommand{\instances}{\mbox{\bf Valuations}} 
\newcommand{\bpar}[1]{\mbox{$\bigl( #1 \bigr)$}}            
\newcommand{\seq}[1]{\langle #1 \rangle}           
\newcommand{\tuple}[2]{\mbox{$ {#1}_{1},\ldots,{#1}_{#2}$}} 
\newcommand{\col}[1]{\mbox{$#1$:}} 
\newcommand{\coli}[1]{\mbox{\scriptsize $#1$:}}

\newcommand{\predell}{\mbox{\bf CL4}}
\newcommand{\propel}{\mbox{\bf CL1}}
\newcommand{\propell}{\mbox{\bf CL2}}
\newcommand{\seqc}{\bigtriangleup}
\newcommand{\seqd}{\bigtriangledown}
\newcommand{\seqst}{\uparrow}
\newcommand{\propelost}{\downarrow}
\newcommand{\mla}{\mbox{{\Large $\wedge$}}}
\newcommand{\mle}{\mbox{{\Large $\vee$}}}
\newcommand{\mlai}{\mbox{{$\wedge$}}}
\newcommand{\mlei}{\mbox{{$\vee$}}}
\newcommand{\gneg}{\neg}                  
\newcommand{\intimpl}{\mbox{\hspace{2pt}$\circ$\hspace{-0.14cm} \raisebox{-0.058cm}{\Large --}\hspace{2pt}}}
\newcommand{\mli}{\rightarrow}                     
\newcommand{\mleq}{\hspace{2pt}\leftrightarrow\hspace{2pt}}   
\newcommand{\cla}{\mbox{\large $\forall$}}      
\newcommand{\cle}{\mbox{\large $\exists$}}        
\newcommand{\mld}{\vee}    
\newcommand{\mlc}{\wedge}  
\newcommand{\ade}{\mbox{\Large $\sqcup$}}      
\newcommand{\ada}{\mbox{\Large $\sqcap$}}      
\newcommand{\add}{\sqcup}                      
\newcommand{\adc}{\sqcap}                      
\newcommand{\clai}{\forall}     
\newcommand{\clei}{\exists}        
\newcommand{\adei}{\mbox{$\sqcup$}}      
\newcommand{\adai}{\mbox{$\sqcap$}}      
\newcommand{\intf}{\$}               
\newcommand{\tlg}{\bot}               
\newcommand{\twg}{\top}               

\newcommand{\st}{\mbox{\raisebox{-0.05cm}{$\circ$}\hspace{-0.13cm}\raisebox{0.16cm}{\tiny $\mid$}\hspace{2pt}}}
\newcommand{\cost}{\mbox{\raisebox{0.12cm}{$\circ$}\hspace{-0.13cm}\raisebox{0.02cm}{\tiny $\mid$}\hspace{2pt}}}
\newcommand{\pst}{\mbox{\raisebox{-0.01cm}{\scriptsize $\wedge$}\hspace{-4pt}\raisebox{0.16cm}{\tiny $\mid$}\hspace{2pt}}}
\newcommand{\pcost}{\mbox{\raisebox{0.12cm}{\scriptsize $\vee$}\hspace{-4pt}\raisebox{0.02cm}{\tiny $\mid$}\hspace{2pt}}}
\newcommand{\ssti}
{\mbox{\raisebox{-0.08cm}
{\large -}\hspace{-0.26cm}
{\tiny $\wedge$}\hspace{-0.135cm}\raisebox{0.08cm}{\tiny $.$}\hspace{-0.079cm}\raisebox{0.10cm}
{\tiny $.$}\hspace{-0.079cm}\raisebox{0.12cm}{\tiny $.$}\hspace{-0.085cm}\raisebox{0.14cm}
{\tiny $.$}\hspace{-0.079cm}\raisebox{0.16cm}{\tiny $.$}\hspace{1pt}}}
\newcommand{\scosti}
{\mbox{\raisebox{0.08cm}
{\large -}\hspace{-0.26cm}
\raisebox{0.08cm}
{\tiny $\vee$}\hspace{-0.135cm}\raisebox{-0.01cm}{\tiny $.$}\hspace{-0.079cm}\raisebox{0.01cm}
{\tiny $.$}\hspace{-0.079cm}\raisebox{0.03cm}{\tiny $.$}\hspace{-0.085cm}\raisebox{0.05cm}
{\tiny $.$}\hspace{-0.079cm}\raisebox{0.07cm}{\tiny $.$}\hspace{1pt}}}

\newcommand{\scc}{\mbox{\scriptsize $\bigtriangleup$}}
\newcommand{\scd}{\mbox{\scriptsize \raisebox{0.04cm}{$\bigtriangledown$}}}
\newcommand{\sca}{\mbox{\small $\bigtriangleup$}}
\newcommand{\sce}{\mbox{\small \raisebox{0.05cm}{$\bigtriangledown$}}}

\newcommand{\sti}{\mbox{\raisebox{-0.02cm}
{\scriptsize $\circ$}\hspace{-0.121cm}\raisebox{0.08cm}{\tiny $.$}\hspace{-0.079cm}\raisebox{0.10cm}
{\tiny $.$}\hspace{-0.079cm}\raisebox{0.12cm}{\tiny $.$}\hspace{-0.085cm}\raisebox{0.14cm}
{\tiny $.$}\hspace{-0.079cm}\raisebox{0.16cm}{\tiny $.$}\hspace{1pt}}}
\newcommand{\costi}{\mbox{\raisebox{0.08cm}
{\scriptsize $\circ$}\hspace{-0.121cm}\raisebox{-0.01cm}{\tiny $.$}\hspace{-0.079cm}\raisebox{0.01cm}
{\tiny $.$}\hspace{-0.079cm}\raisebox{0.03cm}{\tiny $.$}\hspace{-0.085cm}\raisebox{0.05cm}
{\tiny $.$}\hspace{-0.079cm}\raisebox{0.07cm}{\tiny $.$}\hspace{1pt}}}
\newcommand{\psti}
{\mbox{\raisebox{-0.02cm}
{\tiny $\wedge$}\hspace{-0.135cm}\raisebox{0.08cm}{\tiny $.$}\hspace{-0.079cm}\raisebox{0.10cm}
{\tiny $.$}\hspace{-0.079cm}\raisebox{0.12cm}{\tiny $.$}\hspace{-0.085cm}\raisebox{0.14cm}
{\tiny $.$}\hspace{-0.079cm}\raisebox{0.16cm}{\tiny $.$}\hspace{1pt}}}
\newcommand{\pcosti}
{\mbox{\raisebox{0.08cm}
{\tiny $\vee$}\hspace{-0.135cm}\raisebox{-0.01cm}{\tiny $.$}\hspace{-0.079cm}\raisebox{0.01cm}
{\tiny $.$}\hspace{-0.079cm}\raisebox{0.03cm}{\tiny $.$}\hspace{-0.085cm}\raisebox{0.05cm}
{\tiny $.$}\hspace{-0.079cm}\raisebox{0.07cm}{\tiny $.$}\hspace{1pt}}}



\newtheorem{theoremm}{Theorem}[section]
\newtheorem{thesiss}[theoremm]{Thesis}
\newtheorem{definitionn}[theoremm]{Definition}
\newtheorem{figuree}[theoremm]{Figure}
\newtheorem{lemmaa}[theoremm]{Lemma}
\newtheorem{propositionn}[theoremm]{Proposition}
\newtheorem{conventionn}[theoremm]{Convention}
\newtheorem{examplee}[theoremm]{Example}
\newtheorem{remarkk}[theoremm]{Remark}
\newtheorem{conjecturee}[theoremm]{Conjecture}
\newtheorem{claimm}[theoremm]{Claim}
\newtheorem{corollaryy}[theoremm]{Corollary}
\newtheorem{exercisee}[theoremm]{Exercise}

\newenvironment{exercise}{\begin{exercisee} \em}{ \end{exercisee}}
\newenvironment{definition}{\begin{definitionn} \em}{ \end{definitionn}}
\newenvironment{theorem}{\begin{theoremm}}{\end{theoremm}}
\newenvironment{lemma}{\begin{lemmaa}}{\end{lemmaa}}
\newenvironment{corollary}{\begin{corollaryy}}{\end{corollaryy}}
\newenvironment{proposition}{\begin{propositionn} }{\end{propositionn}}
\newenvironment{convention}{\begin{conventionn} \em}{\end{conventionn}}
\newenvironment{remark}{\begin{remarkk} \em}{\end{remarkk}}
\newenvironment{proof}{ {\bf Proof.} }{\  $\Box$ \vspace{.1in} }
\newenvironment{conjecture}{\begin{conjecturee} }{\end{conjecturee}}
\newenvironment{claim}{\begin{claimm} }{\end{claimm}}
\newenvironment{thesis}{\begin{thesiss} }{\end{thesiss}}
\newenvironment{example}{\begin{examplee} \em}{\end{examplee}}
\newenvironment{figuret}{\begin{figuree} \em}{\end{figuree}}

\title{Computability logic: a formal theory of interaction}
\author{Giorgi Japaridze\thanks{This material is based upon work supported by the National Science Foundation under Grant No. 0208816}\\ 
{\small http://www.csc.villanova.edu/$^\sim$japaridz}\\ 
{\small 2004}\\
{\small (written as a book chapter)} 
}
\date{}
\maketitle






\section{Introduction}\label{intr}

In the same sense as classical logic is a formal theory of truth, the recently initiated approach called 
{\em computability logic} (CL) is a formal theory of computability --- in particular, a theory of interactive 
computability. It understands computational problems as games played by a machine against the environment, their computability as existence of a machine that always wins the 
game, logical operators as operations on computational problems, and validity of a logical formula as being a scheme of
``always computable" problems. The paradigm shift in computer science towards interaction provides a solid motivational 
background for CL. In turn, the whole experience of developing CL presents additional strong evidence in favor 
of the new paradigm.  It reveals that the degree of abstraction  
required at the level of logical analysis makes it imperative to understand computability in its most general --- interactive --- sense: the traditional, non-interactive concept of computability appears to be too narrow, and 
its scope delimitation not natural enough, to induce any meaningful logic.

Currently computability logic is at its very first stages of development, with open problems and unverified 
conjectures prevailing over answered questions. A fundamental, 99-page-long introduction to the subject has been given in \cite{Jap03}. The present chapter reintroduces CL in a more compact and less technical way, being written in a 
semitutorial style with a wider computer science audience in mind. 

The traditional Church-Turing approach to computational problems assumes a simple interface between a computing agent and its environment, consisting in asking a question (input) and generating an answer (output). Such an understanding, however, only captures a modest fraction of our broader intuition of computational problems.  This has been not only repeatedly pointed out by the editors and authors of the present collection \cite{Gol00,Gol?,Jap03,Jap04,Mil93,Weg98} but, in fact, 
acknowledged by Turing \cite{Tur36} himself. The reality is that most of the tasks that computers perform are interactive, where not only the computing system but also its environment remain active throughout a computation, with the two parties communicating with each other through what is often referred to as {\em observable actions} \cite{Mil93,Weg98}. Calling 
sequences of observable actions {\em interaction histories}, a computational problem in a broader sense can then be understood as a pair comprising a set of all ``possible" interaction histories and a subset of it of all ``successful" interaction histories; and the computing agent considered to be solving such a problem if it ensures that the actual interaction history is always among the successful ones. 

As this was mentioned, technically CL understands interactive problems as {\em games}, or {\em dialogues}, between two agents/players:  {\em machine} and  {\em environment}, symbolically named as $\pp$ and $\oo$\label{y90}, respectively. Machine, as its name suggests, is specified as a mechanical 
device with fully determined, effective behavior, while the behavior of the environment, which represents a capricious 
user or the blind forces of nature, is allowed to be arbitrary. Observable actions by these two agents translate into 
game-theoretic terms as their {\em moves}, interaction histories as {\em runs}, i.e. sequences of moves, ``possible" interaction histories as {\em legal runs}, and ``successful" interaction histories as {\em won {\em (by $\pp$)} runs}. 

Computational problems in the Church-Turing sense are nothing but dialogues/games of depth 2, with the first legal move (``input") by the 
environment and the second legal move (``output") by the machine. The problem of finding the value of a function $f$
is a typical task modeled by this sort of games. In the formalism of CL this problem is expressed by the formula\vspace{-3pt} \[\ada x\ade y\bigl(y=f(x)\bigr).\vspace{-3pt}\] It stands for a two-move-deep game where the first move --- selecting a particular value $m$ for $x$ --- must be made by $\oo$, and the second move --- selecting a value $n$ for $y$ --- by $\pp$. The game is then considered won by the machine, i.e., the problem solved, if $n$ really equals $f(m)$. So,  computability of $f$ means nothing but existence of a machine that wins the game $\ada x\ade y\bigl(y=f(x)\bigr)$ against any possible (behavior of the) environment. 

Generally, $\ada xA(x)$\label{0ada0} is a game where the environment has to make the first move by selecting a particular value $m$ for $x$, after which the play continues --- and the winner is determined --- according to the rules of $A(m)$; if $\oo$ fails to make an initial move, the game is considered won by the machine as there was no particular (sub)problem specified by its adversary that it failed to solve. $\ade xA(x)$\label{0ade0} is defined in the same way, only here it is the machine who makes an initial move/choice and it is the environment who is considered the winner if such a choice is never made. This interpretation makes $\ade$ a constructive version of existential quantifier, while $\ada$ is a constructive version of universal quantifier. 

As for standard atomic formulas, such as $n=f(m)$, they are understood as games without any moves. This sort of games are called {\em elementary}.\label{0elg} An elementary game is automatically won or lost by the machine depending on whether the formula representing it is true or false (true=won, false=lost). This interpretation makes the classical concept of predicates\label{0predicate0}  a special case of games.

The meanings of the propositional counterparts $\add$\label{0adc0} and $\adc$\label{0add0} of $\ade$ and $\ada$ 
are not hard to guess. They, too, signify a choice by the corresponding player. The only difference is that while in the case of $\ade$ and $\ada$ the choice is made among the objects of the universe of discourse, $\add$ and $\adc$ mean a choice between {\em left} and {\em right}. For example, the problem of deciding predicate $P(x)$ could be expressed 
by \(\ada x\bigl(P(x) \add \gneg P(x)\bigr),\) denoting the game where the environment has to select a value $m$ for $x$,
to which the machine should reply by one of the moves {\em left {\em or} right}; the game will be considered won by the machine if $P(m)$ is true and the move {\em left} was made or $P(m)$ is false and the choice was {\em right}, so that decidability of $P(x)$ means nothing but existence of a machine that always wins the game $\ada x\bigl(P(x) \add 
\gneg P(x)\bigr)$. 

The above example involved classical negation $\gneg$. The other classical operators will also be allowed in our language, and they all acquire a new, natural game interpretation. The reason why we can still call them ``classical" is that, when applied to elementary games --- i.e. predicates --- they preserve the elementary property of their arguments and act exactly in the classical way. Here is an informal explanation of how the ``classical" operators are understood as game operations: 

The game $\gneg A$\label{0gneg0} is nothing but $A$ with the roles of the two players switched: $\pp$'s moves or wins become $\oo$'s moves or wins, and vice versa.  For example, where {\em Chess} is the game of chess (with the possibility of draw outcomes ruled out for simplicity) from the point of view of the white player,  $\gneg \mbox{\em Chess}$ is the same game from the point of view of the black player. 

The operations $\mlc$\label{0mlc0} and $\mld$\label{0mld0} combine games in a way that corresponds to the intuition of parallel computations. Playing $A\mlc B$ or $A\mld B$ means playing the two games $A$ and $B$ simultaneously. In $A\mlc B$ the machine is considered the winner if it wins in both of the components, while in $A\mld B$ it is sufficient to win in one of the components. 
Thus we have two sorts of conjunction: $\adc,\mlc$ and two sorts of disjunction: $\add,\mld$. Comparing the games ${\em Chess}\mld\gneg \mbox{\em Chess}$ and $\mbox{\em Chess} \add \gneg \mbox{\em Chess}$ will help us appreciate the difference. 
The former  is, in fact, a parallel play on two boards, where $\pp$ plays white on the left board and black on the right board. There is a strategy for $\pp$ that guarantees an easy success in this game even if the adversary is a world champion. All that $\pp$ needs to do is to mimic, in {\em Chess}, the moves made by $\oo$ in $\gneg \mbox{\em Chess}$, and vice versa. On the other hand, winning the game $\mbox{\em Chess}\hspace{0.02in} \add\gneg \mbox{\em Chess}$ is not easy at all: here, at the very beginning, $\pp$ has to choose between {\em Chess} and $\gneg \mbox{\em Chess}$ and then win the chosen one-board game. Generally, the principle $A\mld\gneg A$ is valid in the sense that the corresponding problem is always solvable by a machine, whereas this is not so for $A\add\gneg A$. 

While all the classical tautologies automatically hold when classical operators are applied to elementary games, in the general case the class of valid principles shrinks. For example,  $\gneg A\mld (A\mlc A)$ is not valid. The above ``mimicking strategy" would obviously be inapplicable in the three-board game \(\gneg \mbox{\em Chess}\mld (\mbox{\em Chess}\mlc \mbox{\em Chess})\): the best that $\pp$ can do here is to pair $\gneg \mbox{\em Chess}$ with one of the two conjuncts of $\mbox{\em Chess}\mlc \mbox{\em Chess}$. It is possible that then $\gneg \mbox{\em Chess}$ and the unmatched {\em Chess} are both lost, in which case the whole game will be lost.  

The class of valid principles of computability forms a logic that resembles linear logic\label{ll1} \cite{Gir87} with $\gneg$ understood as linear negation, $\mlc,\mld$ as multiplicatives and $\adc,\add,\ada,\ade$ as additives. It should be however pointed out that, despite similarity, these computability-logic operations are by no means ``the same" as those of linear logic (see Section \ref{sull}). To stress the difference and avoid possible confusion, we refrain from using any linear-logic terminology, calling $\adc,\add,\ada,\ade$ {\em choice operations} and $\mlc,\mld$\label{0parop} and {\em parallel operations}. 

Assuming that the universe of discourse is $\{1,2,3,\ldots\}$, obviously the meanings of $\ada xA(x)$ and $\ade xA(x)$ can be explained as $A(1)\adc A(2)\adc A(3)\adc\ldots$ and $A(1)\add A(2)\add A(3)\add\ldots$, respectively. Similarly, our parallel operations $\mlc$ and $\mld$ have their natural quantifier-level counterparts $\mla$ and $\mle$, with 
$\mla xA(x)$ understood as $A(1)\mlc A(2)\mlc A(3)\mlc\ldots$ and $\mle xA(x)$ as $A(1)\mld A(2)\mld A(3)\mld\ldots$. 
Hence, just like $\mlc$ and $\mld$, the operations $\mla$ and $\mle$ are ``classical" in the sense that, when applied to elementary games, they  behave exactly as the classical universal and existential quantifiers, respectively. 

The {\em parallel implication} $\mli$\label{0mli0} --- yet another ``classical" operation --- is perhaps most  interesting from the 
computability-theoretic point of view. Formally $A\mli B$ is defined as $\gneg A\mld B$.  The intuitive meaning of $A\mli B$ is the problem of {\em reducing} problem $B$ to problem $A$. Putting it in other words, solving $A\mli B$ means solving $B$ having $A$ as an (external) {\em resource}.\label{0cr} ``Resource" is symmetric to ``problem": what is a problem (task) for the machine, is a resource for the environment, and vice versa.
To get a feel of $\mli$ as a problem reduction operator, consider the reduction of the acceptance problem to the halting problem. The halting problem  can be expressed by\vspace{-3pt} 
\[\ada x\ada y \bigl(\mbox{\em Halts}(x,y) \add \gneg \mbox{\em Halts}(x,y)\bigr),\vspace{-3pt}\]
 where $\mbox{\em Halts}(x,y)$ is the predicate  
``Turing machine  $x$ halts on input $y$". And the acceptance problem can be expressed by\vspace{-3pt} 
\[\ada x\ada y \bigl(\mbox{\em Accepts}(x,y) \add \gneg \mbox{\em Accepts}(x,y)\bigr),\vspace{-3pt}\] 
with $\mbox{\em Accepts}(x,y)$ meaning 
``Turing machine  $x$ accepts input $y$". While the acceptance problem is not decidable, it is effectively reducible to the halting problem. In particular, there is a machine that always wins the game\vspace{-3pt}
\[
\ada x\ada y \bigl(\mbox{\em Halts}(x,y)\add \gneg \mbox{\em Halts}(x,y)\bigr) \mli\ \ada x\ada y \bigl(\mbox{\em Accepts}(x,y)\add \gneg \mbox{\em Accepts}(x,y)\bigr).\vspace{-3pt}
\]
A strategy for solving this problem is to wait till $\oo$ specifies values $m$ and $n$ for $x$ and $y$ in the consequent, then select the same values $m$ and $n$ for  $x$ and $y$ in the antecedent (where the roles of $\pp$ and $\oo$ 
are switched), and see whether $\oo$ responds by {\em left} or {\em right} there. If the response is {\em left}, simulate machine $m$ on input $n$ until it halts and then select, in the consequent, {\em left} or {\em right} depending on whether the simulation accepted or rejected. And if $\oo$'s response in the antecedent was {\em right}, then select {\em right} in the consequent. 

What the machine did in the above strategy was indeed a reduction of the acceptance problem to the halting problem: 
it solved the former by employing an external, environment-provided solution to the latter. A strong case can be made in favor of the thesis that $\mli$ captures our ultimate intuition of reducing one interactive problem to another. It should be noted, however, that the reduction captured by $\mli$ is stronger than Turing reduction, which is often perceived as
an adequate formalization of our most general intuition of reduction. Indeed, if we talk in terms of oracles that the definition of Turing reduction employs, specifying the values of $x$ and $y$ as $m$ and $n$ in the antecedent can be thought of as asking the oracle regarding whether machine $m$ halts on input $n$.   
Notice, however, that the usage of the oracle here is limited as it can only be employed once: after querying regarding 
$m$ and $n$, the machine would not be able to repeat the same query with different parameters $m'$ and $n'$, for that would require having two ``copies" of the resource 
$\ada x\ada y \bigl(\mbox{\em Halts}(x,y) \add \gneg \mbox{\em Halts}(x,y)\bigr)$ (which could be expressed by their $\mlc$-conjunction) rather than one. On the other hand, Turing reduction allows recurring usage of the oracle, which 
the resource-conscious CL understands as reduction not to the halting problem $\ada x\ada y \bigl(\mbox{\em Halts}(x,y)\add \gneg \mbox{\em Halts}(x,y)\bigr)$ but to the stronger problem expressed by $\pst\ada x\ada y \bigl(\mbox{\em Halts}(x,y)\add \gneg \mbox{\em Halts}(x,y)\bigr)$. Here $\pst A$, called the {\em parallel recurrence}
of $A$, means the infinite conjunction $A\mlc A\mlc\ldots$. The $\pst$-prefixed halting problem
now explicitly allows an unbounded number of queries of the type ``does $m$ halt on $n$?". So, Turing reducibility of $B$ to $A$ --- which, of course, is only defined when $A$ and $B$ are computational problems in the traditional sense, i.e. problems of the type
$\ada x\bigl(\mbox{\em Predicate}(x)\add\gneg \mbox{\em Predicate}(x)\bigr)$ 
or $\ada x\ade y\mbox{\em Predicate}(x,y)$ --- 
means computability of $\pst A\mli B$ 
rather than $A\mli B$, i.e. reducibility of $B$ to $\pst A$ rather than to $A$.
To put this intuition together, consider the Kolmogorov complexity problem.\label{kc} It can be expressed by
\(\ada t\ade z\hspace{2pt}\mbox{\em K}\hspace{2pt}(z,t),\)
where  $\mbox{\em K}\hspace{2pt}(z,t)$ is the predicate ``$z$ is the smallest (code of a) Turing machine that returns $t$ on input $1$".  Having no algorithmic solution, the Kolmogorov complexity problem, however, is known to be Turing reducible to the halting problem. In our terms, this means nothing but that there is a machine that always wins the game\vspace{-3pt}
\begin{equation}\label{kce}\pst\ada x\ada y \bigl(\mbox{\em Halts}(x,y) \add \gneg {\em Halts}(x,y)\bigr)\ \mli\ \ada t\ade z\hspace{2pt} \mbox{\em K}\hspace{2pt}(z,t).\vspace{-3pt}\end{equation}
Here is a strategy for such a machine: Wait till $\oo$ selects a value $m$ for $t$  in the consequent. 
Then, starting from $i=1$, do the following: in the $i$th $\mlc$-conjunct  of the antecedent, make two consecutive  moves
 by specifying $x$ and $y$ as $i$ and $1$, respectively. If $\oo$ responds there by {\em right},
increment $i$ by one and repeat the step; if $\oo$ responds by {\em left}, simulate machine $i$ on input $1$ until it halts; if you see that  machine $i$ returned $m$, make a move in the consequent by specifying $z$ as $i$; otherwise, increment $i$ by one and repeat the step.

One can show that Turing reduction of the Kolmogorov complexity problem to the halting problem essentially requires unlimited usage of the oracle, which means that, unlike the acceptance problem, the Kolmogorov complexity problem is not reducible to the halting problem in our sense, and is only reducible to the parallel recurrence of it. 
That is, (\ref{kce}) is computable but not with $\pst$ removed from the antecedent. One might expect that $\pst A\mli B$ captures the intuition of reducing an interactive problem $B$ to an interactive problem $A$ in the {\em weakest} sense, as Turing reduction certainly does for non-interactive problems. But this is not so. While having $\pst A$ in the antecedent,
i.e. having it as a resource, indeed allows an agent to reuse $A$ as many times as it wishes, there is a stronger --- in fact the strongest --- form of reusage, captured by another operation $\st$ called {\em branching recurrence}. Both $\pst A$ and 
$\st A$ can be 
thought of as games where $\oo$ can restart $A$ as many times as it likes. The difference is that in $\st A$, unlike $\pst A$,  $A$ can be restarted not only from the very beginning, but from any already reached state. 
This gives $\oo$ greater flexibility, such as, say, the capability to try different answers to the same (counter)question  by $\pp$, while this could be impossible in $\pst A$ because $\pp$ may have asked different questions in different conjuncts of $\pst A$. Section 2 will explain the differences between $\pst$ and $\st$ in more detail. Our claim that $\st$ captures the strongest sort of resource-reusage automatically translates into another claim, according to which $\st A\mli B$ captures the weakest possible sort of reduction of one interactive problem to another. The difference between $\st$ and $\pst$ is
irrelevant when they are applied to two-step ``traditional" problems such as the halting problem or the Kolmogorov complexity problem: for such a ``traditional" problem $A$, $\pst A$
and $\st A$ turn out to be logically equivalent, and hence both $\pst A\mli B$ and $\st A\mli B$ are equally accurate translations for Turing reduction of $B$ to $A$. The equivalence between $\st$ and $\pst$, however, certainly does not extend to the general case. E.g., the principle $\st(A\add B)\mli \st A\add\st B$ is valid while    $\pst(A\add B)\mli \pst A\add\pst B$ is not. Among the so far unverified conjectures of CL is that the logical behavior of $\st A\mli  B$ is exactly that of the implication $A\intimpl B$ of (Heyting's) intuitionistic logic. 

Another group of operations that play an important role in CL comprises  $\cla$\label{0cla0} and its dual $\cle$\label{0cle0} (with $\cle xA(x)=\gneg\cla x\gneg A(x)$), called  {\em blind quantifiers}.\label{0bq}  $\cla xA(x)$ 
can be thought of as a ``version" of $\ada xA(x)$ where the particular value of $x$ that the environment selects is invisible to the machine, so that it has to play blindly in a way that guarantees success no matter what that value is. This way, $\cla$ and $\cle$  produce games with {\em imperfect information}.  

Compare the problems
\(\ada x\bigl(\mbox{\em Even$(x)$}\add \mbox{\em Odd$(x)$}\bigr)\) and  \(\cla x\bigl(\mbox{\em Even$(x)$}\add \mbox{\em Odd$(x)$}\bigr).\) 
Both of them are about telling whether a given number is even or odd; the difference is only in whether that ``given number" is known to the machine or not. The first problem is an easy-to-win, two-move-deep game of a structure that we have already seen.  The second game, on the other hand, is one-move deep with only by the machine to make a move --- select the ``true"  disjunct, which is hardly possible to do as the value of $x$ remains unspecified. 

Of course, not all nonelementary $\cla$-problems will be unsolvable. Here is an example:\vspace{-3pt}
\[\cla x\Bigl(\mbox{\em Even$(x)$}\add \mbox{\em Odd$(x)$}\ \mli\ \ada y\bigl(\mbox{\em Even$(x\times y)$}\add
\mbox{\em Odd$(x\times y)$}\bigr)\Bigr).\vspace{-3pt}\]
Solving this problem, which means reducing the consequent to the antecedent without knowing the value of $x$, is easy: 
$\pp$ waits till $\oo$ selects  a value $n$ for $y$. If $n$ is even, then $\pp$ makes the move {\em left} in the consequent. Otherwise, if $n$ is odd, $\pp$ continues waiting until $\oo$ selects one of the $\add$-disjuncts in the antecedent
(if $\oo$ has not already done so), and then $\pp$ makes the same move 
{\em left}  or {\em right} in the consequent as $\oo$ made in the antecedent. Note that our semantics for $\ada,\add,\mli$ 
guarantees an automatic win for $\pp$ if $\oo$ fails to make either selection.

Both $\cla xA(x)$ and $\mla xA(x)$ can be shown to be properly stronger that $\ada xA(x)$, in the sense that 
$\cla xA(x)\mli \ada xA(x)$ and $\mla xA(x)\mli\ada xA(x)$ are valid while  $\ada xA(x)\mli \cla xA(x)$ and $\ada xA(x)\mli\mla xA(x)$ are not. On the other hand, the strengths of $\cla xA(x)$ and $\mla xA(x)$ are mutually incomparable: 
neither $\cla xA(x)\mli \mla x A(x)$ nor $\mla xA(x)\mli \cla x A(x)$ is valid. 
The big difference between $\cla$ and $\mla$ is that, while playing $\cla xA(x)$ means playing one ``common" play for 
all possible $A(c)$ and thus $\cla xA(x)$ is a  
one-board game, $\mla xA(x)$ is an infinitely-many-board game: playing it means playing, in parallel, 
game $A(1)$ on board \#1, game $A(2)$ on board \#2, etc.
When restricted to elementary games, however, the distinction between the blind and the parallel groups of quantifiers disappears and, just like $\gneg$, $\mlc$, $\mld$, $\mli$, $\mla$, $\mle$, the blind quantifiers behave exactly in the classical way. Having this collection of operators makes CL a conservative extension of classical first-order logic: the latter is nothing but CL restricted to elementary problems and the logical vocabulary $\gneg$, $\mlc$, $\mld$, $\mli$, $\cla$ (and/or $\mla$), $\cle$ (and/or $\mle$). 

 As the above examples illustrate, what can be considered an adequate formal equivalent of our broad intuition of computational problems goes far beyond the traditional, two-step, input/output problems. Computational problems of higher degrees of interactivity emerge naturally and have to be addressed in any more or less advanced study in computability theory. So far this has been  mostly done in an ad hoc manner as there 
has been no standard way for specifying interactive problems. The formalism of CL offers a convenient language for expressing interactive computational problems and studying them in a systematic way. Finding effective axiomatizations of the corresponding logic or, at least, some reasonably rich fragments of it, is expected to have not only  theoretical,  but also high practical significance. Among the applications would be  the possibility to build CL into a machine and then use such a machine as a universal problem-solving tool. 

Outlining the rest of this chapter: Sections 2-4 provide formal definitions --- accompanied with explanations and illustrations --- of the basic 
concepts on interactive computational problems understood as games, including the main operations on such problems/games.
Section 5 introduces a model of interactive computation that generalizes Turing machines and, allowing us to extend the Church-Turing thesis to interaction, serves as a basis for our definition of interactive computability. 
Sections 6 and 7 present sound and complete axiomatizations of various fragments of CL. Section 8 discusses, using numerous examples and illustrations, potential applications 
of CL  in the areas of (interactive) knowledgebase systems, planning systems and constructive applied theories.

\section{Constant games}\label{wig}

Our ultimate concept of games will be defined in the next section in terms of the simpler and more basic class of games 
called constant games. 
To define this class, we need some technical terms and conventions. Let us agree that by a {\bf move}\label{y31} we mean any finite string over the standard keyboard alphabet. 
A {\bf labeled move}\label{x39} is a move prefixed with $\pp$ or $\oo$, with its prefix ({\bf label}) indicating which player has made the move. 
A {\bf run}\label{y41} is a (finite or infinite) sequence of labeled moves, and a {\bf position}\label{y40} is a finite run. 

We will be exclusively using letters $\Gamma,\Upsilon$ as metavariables for runs, $\Phi,\Psi$  for positions, $\xx$ for players, and $\alpha$ for moves.
Runs will be often delimited with ``$\langle$" and ``$\rangle$", with $\emptyrun$\label{yy3} thus denoting the {\bf empty run}. The meaning of an expression such as $\seq{\Phi,\xx\alpha,\Gamma}$ must be clear: this is the result of appending to $\seq{\Phi}$ 
 $\seq{\xx\alpha}$ and then $\seq{\Gamma}$.

\begin{definition}\label{game}
 A {\bf constant game} is a pair $A=(\legal{A}{},\win{A}{})$, where:

1. $\legal{A}{}$ is a set of runs satisfying the condition\footnote{\cite{Jap03}\label{ftn} imposes an additional condition according 
to which there is a special move that no element of $\legal{A}{}$ contains. The only result of 
\cite{Jap03} that appeals to that condition is Lemma 4.7. In the present exposition we directly incorporate the statement 
of that lemma into the definition of static games (page \pageref{stg}), and thus all results of \cite{Jap03} --- in particular, those that 
rely on Lemma 4.7 --- remain valid. This and a couple of other minor technical differences between our present formulations
 from those given in \cite{Jap03} only signify presentational and by no means conceptual variations.}
that a (finite or infinite) run $\Gamma$ is in $\legal{A}{}$ iff all of its nonempty finite initial segments are in $\legal{A}{}$.

2. $\win{A}{}$  is a function of the type $\legal{A}{}\rightarrow\{\pp,\oo\}$. We use $\win{A}{}\seq{\Gamma}$ 
to denote the value of $\win{A}{}$ at $\Gamma$.
\end{definition}

The intuitive meaning of the $\legal{A}{}$ component of a constant game $A$, called the {\bf structure of $A$}, 
is that it tells us what runs are legal. Correspondingly, we call the elements of $\legal{A}{}$ the {\bf legal runs} of $A$, and call all other runs {\bf illegal}. For a player $\xx$, a run $\Gamma$ is said to be a {\bf $\xx$-legal}\label{lr} run of $A$ iff either $\Gamma$ is a legal run of $A$ or otherwise the label of the last move of 
the shortest illegal initial segment of $\Gamma$ is not $\xx$. 
Understanding an {\em illegal move}\label{y19} by player $\xx$ in position $\Phi$ as a move $\alpha$ such that adding $\xx\alpha$ 
to $\Phi$ makes this position illegal, the condition of clause 1 of Definition \ref{game} corresponds to the intuition that a run is legal iff no illegal moves have been made in it, which automatically implies that the empty position $\emptyrun$ is a legal run of every game.  And a $\xx$-legal run of $A$ is a run where $\xx$ has not made any illegal moves in any of the legal positions --- in other words, a run where, if there are illegal moves at all, $\xx$ is not the first to have made 
such a move.    
 When modeling real-life interactive tasks, such as server-client or robot-environment interaction, illegal moves 
will usually mean actions that can, will or should never be performed.  For generality, flexibility and convenience, 
our approach however does not exclude illegal runs from considerations.  
  
As for the 
$\win{A}{}$ component of a constant game $A$, called the {\bf content} of the game, it tells us who has won 
a given legal run. A run $\Gamma\in\legal{A}{}$ with $\win{A}{}\seq{\Gamma}=\xx$ will be said to be a {\bf $\xx$-won} run of $A$.
 
We say that a constant game $A$ is {\bf elementary}\label{y14} iff $\legal{A}{}=\{\emptyrun\}$. Thus, elementary games  have no legal moves:
the empty run $\emptyrun$ is the only legal run of such games. There are exactly two elementary constant games, for which we use the same symbols $\pp$ and $\oo$ as for the two players. They are defined by stipulating that 
($\legal{\twg}{}=\legal{\tlg}{}=\{\emptyrun\}$ and) $\win{\twg}{}\emptyrun=\pp$, $\win{\tlg}{}\emptyrun=\oo$. Below comes an official definition of some of the basic game operations informally explained in Section 1.

\begin{definition}\label{op}
 In each of the following clauses, $\Phi$ ranges over {\em nonempty} positions --- in view of Definition \ref{game}, it would be sufficient to define $\legal{}{}$ only for this sort of $\Phi$, for then $\legal{}{}$ uniquely extend to all runs.
$\Gamma$ ranges over the legal runs of the game that is being defined.  $A,A_1,A_2$ are any constant games. The notation $\bar{\Phi}$\label{yy11} in clause 1 means 
the result of interchanging $\pp$ with $\oo$ in all labeled moves of $\Phi$. 
And the notation 
$\Phi^{i.}$ in clauses 2 and 3 means the result of removing from $\Phi$ all labeled moves 
except those of the form $\xx i.\alpha$ ($\xx\in\{\pp,\oo\}$), and then deleting the prefix `$i.$' in the remaining moves, i.e. replacing each such $\xx i.\alpha$ by $\xx\alpha$. Similarly for $\bar{\Gamma}$, $\Gamma^{i.}$.
\begin{enumerate}
\item {\bf Negation}\label{y32} $\gneg A$:
\begin{itemize}
\item $\Phi\in\legal{\gneg A}{}$ iff $\bar{\Phi}\in\legal{A}{}$;
\item $\win{\gneg A}{}\seq{\Gamma}=\pp$ iff $\win{A}{}\seq{\bar{\Gamma}}=\oo$.
\end{itemize}
\item {\bf Parallel conjunction}\label{y34} $A_1\mlc A_2$:
\begin{itemize}
\item $\Phi\in \legal{A_1\mlc A_2}{}$ iff every move of $\Phi$ starts with `$1.$' or `$2.$' and,
for each $i\in\{1,2\}$,   $\Phi^{i.}\in\legal{A_i}{}$; 
\item $\win{A_1\mlc A_2}{}\seq{\Gamma}=\pp$ iff, for each $i\in\{1,2\}$,     
$\win{A_i}{}\seq{\Gamma^{i.}}=\pp$.
\end{itemize}
\item {\bf Parallel disjunction}\label{y35} $A_1\mld A_2$:
\begin{itemize}
\item $\Phi\in \legal{A_1\mld A_2}{e}$ iff every move of $\Phi$ starts with `$1.$' or `$2.$' and,
for each $i\in\{1,2\}$,  $\Phi^{i.}\in\legal{A_i}{}$; 
\item $\win{A_1\mld A_2}{}\seq{\Gamma}=\oo$ iff, for each $i\in\{1,2\}$, $\win{A_i}{}\seq{\Gamma^{i.}}=\oo$.
\end{itemize}
\item {\bf Choice conjunction} $A_1\adc  A_2$:\label{y2}
\begin{itemize}
\item $\Phi\in\legal{A_1\adc A_2}{}$ iff $\Phi=\seq{\oo i,\Psi}$, where $i\in\{1,2\}$ and 
$\Psi\in\legal{A_i}{}$; 
\item $\win{A_1\adc A_2}{}\seq{\Gamma}=\oo$ iff $\Gamma=\seq{\oo i,\Upsilon}$, where $i\in\{1,2\}$ and $\win{A_i}{}\seq{\Upsilon}=\oo$.
\end{itemize}
\item {\bf Choice disjunction} $A_1\add A_2$:\label{yy1}
\begin{itemize}
\item $\Phi\in\legal{A_1\add A_2}{}$ iff $\Phi=\seq{\pp i,\Psi}$, where $i\in\{1,2\}$ and 
$\Psi\in\legal{A_i}{}$; 
\item $\win{A_1\add A_2}{}\seq{\Gamma}=\pp$ iff $\Gamma=\seq{\pp i,\Upsilon}$, where $i\in\{1,2\}$ and $\win{A_i}{}\seq{\Upsilon}=\pp$.
\end{itemize}
\item {\bf Parallel implication}, or {\bf reduction} $A_1\mli A_2$\label{yy0} is defined as $(\gneg A_1)\mld A_2$.  
\end{enumerate}

\end{definition}

The operations $\mlc,\mld,\adc,\add$ naturally generalize from binary to $n$-ary (any natural number $n$) or even infinite-ary, where the $0$-ary $\mlc$ and $\adc$ should be understood as $\twg$ and the $0$-ary $\mld$ and $\add$ as $\tlg$. Alternatively, $A_1\mlc\ldots\mlc A_n$ with $n>2$ can be understood as an abbreviation for $A_1\mlc(A_2\mlc\ldots (A_{n-1}\mlc A_n)\ldots)$. Similarly for $\mld,\adc,\add$. For simplicity, officially we will stick 
to the binary version. 

Notice the perfect symmetry/duality between $\mlc$ and $\mld$ or  $\adc$ and $\add$: the definition of 
each of these operations can be obtained from the definition of its dual by interchanging $\pp$ with $\oo$. We earlier 
characterized legal plays of $A_1\mlc A_2$ and $A_1\mld A_2$ as plays ``on two boards". According to the above definition, making 
a move $\alpha$ on ``board" $\# i$ is technically done by prefixing $\alpha$ with `$i.$'.

\begin{exercise}\label{ex01}
 Verify the following equalities (any constant games $A$,$B$):\vspace{3pt}

$\begin{array}{l}
1.\ \ \tlg=\gneg\twg; \ \ \twg=\gneg \tlg;\\
2.\ \ A = \gneg\gneg A;\\
3.\ \ A\mlc B = \gneg(\gneg A\mld\gneg B); \ \ A\mld B = \gneg(\gneg A\mlc\gneg B);\\
4.\ \ A\adc B = \gneg(\gneg A\add\gneg B); \ \ A\add B = \gneg(\gneg A\adc\gneg B).\vspace{0pt}
\end{array}$
\end{exercise}

\begin{exercise}\label{ex08}
Verify that both $\seq{\oo 1.1,\pp 2.1.2}$ and $\seq{\pp 2.1.2,\oo 1.1}$ are (legal and) $\pp$-won runs of the game
$(\twg\add\tlg)\mli \bigl((\tlg\add \twg)\mlc\twg\bigr)$, i.e. --- by Exercise \ref{ex01} --- of the game $(\tlg\adc \twg)\mld \bigl((\tlg\add \twg)\mlc\twg\bigr)$. How about the runs $\emptyrun$, $\seq{\oo 1.1}$, $\seq{\pp 2.1.2}$?
\end{exercise}

An important game operation not mentioned in Section 1 is that of prefixation, which is somewhat reminiscent of 
the modal operator(s) of dynamic logic. This operation takes two arguments: a constant game $A$ and a position $\Phi$ 
 that must 
be a legal position of $A$ (otherwise the operation is undefined). 

\begin{definition}\label{prfx}
Let $A$ be a constant game and $\Phi$ a legal position of $A$. The \mbox{\bf $\Phi$-prefixation} of $A$, denoted $\seq{\Phi}A$, is defined as follows:\vspace{-2pt}
\begin{itemize}
\item $\legal{\seq{\Phi}A}{}=\{\Gamma\ |\ \seq{\Phi,\Gamma}\in\legal{A}{}\}$.\vspace{-2pt}
\item $\win{\seq{\Phi}A}{}\seq{\Gamma}=\win{A}{}\seq{\Phi,\Gamma}$ \ (any $\Gamma\in\legal{\seq{\Phi}A}{}$).
\end{itemize} 
\end{definition}
  
Intuitively, $\seq{\Phi}A$ is the game playing which means playing $A$ starting (continuing) from position $\Phi$. 
That is, $\seq{\Phi}A$ is the game to which $A$ evolves (will be ``{\em brought down}") after the moves of $\Phi$ have been made. We have already used this intuition when explaining the meaning of choice operations: we said that after $\oo$ makes an initial move $i\in\{1,2\}$,
 the game 
$A_1\adc A_2$ continues as $A_i$. What this meant was nothing but that 
$\seq{\oo i}(A_1\adc A_2)=A_i$.
Similarly, $\seq{\pp i}(A_1\add A_2)=A_i$.

\begin{exercise}\label{mar12}
 Verify that, for arbitrary constant games $A,B$, we have:

1. Where $\seq{\xx_1\alpha_1,\ldots,\xx_n\alpha_n}\in\legal{A}{}$,  \ $\seq{\xx_1\alpha_1,\ldots,\xx_n\alpha_n}A=
\seq{\xx_n\alpha_n}\ldots\seq{\xx_1\alpha_1}A$. 

2. Where $\seq{\pp\alpha}\in\legal{\gneg A}{}$, \ $\seq{\pp\alpha}\gneg A=\gneg\seq{\oo\alpha} A$. Same with $\twg,\tlg$ interchanged. 

3. Where $\seq{\xx 1.\alpha}\in \legal{A\mlc B}{}$, \ 
  $\seq{\xx 1.\alpha}(A\mlc B)=(\seq{\xx\alpha}A)\mlc B$. Similarly for $\seq{\xx 2.\alpha}$. Similarly for $A\mld B$.  
\end{exercise}

Prefixation is very handy in visualizing legal runs of a given game $A$. In particular, every (sub)position $\Phi$ of such a run  can be represented by, or thought of as, the game $\seq{\Phi}A$. 

\begin{example}\label{nov26}
Let $G_0=\bigl(A\adc\ (B\add C)\bigr)\mlc \bigl(D\mld(E\add F)\bigr)$.  
The run $\seq{\pp 2.2.1,\oo 1.2,\pp 1.2}$ 
is a legal run of $G_0$, and to it 
corresponds the following sequence of games:\vspace{3pt}





$\begin{array}{lllll}
G_0: & \bigl(A\adc (B\add C)\bigr) \mlc \bigl(D\mld( E\add F)\bigr), & & \mbox{i.e. $G_0$,} & \mbox{i.e. $\emptyrun G_0$;} \\

G_1: & \bigl(A\adc (B\add C)\bigr) \mlc (D\mld E), &  & \mbox{i.e. $\seq{\pp 2.2.1}G_0$,} & \mbox{i.e. $\seq{\pp 2.2.1}G_0$;}\\ 

G_2: & (B\add C) \mlc(D\mld E), &  &   \mbox{i.e. $\seq{\oo 1.2}G_1$,} & \mbox{i.e. $\seq{\pp 2.2.1,\oo 1.2}G_0$;}\\ 

G_3: &  C \mlc (D\mld E), &  & \mbox{i.e. $\seq{\pp 1.2}G_2$,} & \mbox{i.e. $\seq{\pp 2.2.1,\oo 1.2,\pp 1.2}G_0$.}\vspace{2pt} 
\end{array}$

The run stops at $C\mlc (D\mld E)$, and hence the winner is the player $\xx$ with 
$\win{C\mlc (D\mld E)}{}\emptyrun=\xx$. Note how the $\mlc,\mld$-structure of the game was retained throughout the play.   \end{example}

Another constant-game operation of high interest is {\bf branching recurrence} $\st$. A strict formal definition of this operation, together with 
detailed discussions and illustrations of the associated intuitions, can be found in Section 13 of \cite{Jap03}.
Here we only give a brief informal explanation. A legal run of $\st A$ can be thought of as a tree rather than sequence of labeled moves (with those labeled moves associated with the edges --- rather than nodes --- of the tree), where each 
branch of the tree spells a legal run of $A$.  $\pp$ is considered the winner in such a game iff it wins $A$ in {\em all} of the branches. The play starts with the root as the only node of the tree, representing the empty run; at any time, $\pp$ can make any legal move of $A$ in any of the existing branches. 
So can $\oo$, with the difference that $\oo$ --- and only $\oo$ --- also has the capability, by making a special ``splitting" move (that we do not count as a move of $A$), to fork any given branch into two, thus creating two runs of $A$ out of one that share the same beginning but from now on can evolve in different ways. So, $\st$ allows $\oo$ to replicate/restart $A$ as many times as it wishes; furthermore, as noted in Section 1, $\oo$ does not really have to restart $A$ from the very beginning every time it ``restarts" it; instead, $\oo$ may choose to continue a new run of $A$ from any already reached position $\Phi$ of $A$, i.e. replicate $\seq{\Phi}A$ rather than $A$,  thus depriving $\pp$ of the possibility to  reconsider its previously 
made moves while giving itself the opportunity to try different strategies in different 
continuations of $\Phi$ and become the winner as long as one of those strategies succeeds. This makes $\st A$ easier for $\oo$ to win than the infinite conjunction $A\mlc A\mlc A\mlc\ldots$ that we call {\bf parallel recurrence} $\pst A$. The latter 
can be considered a restricted version of $\st A$ where all the branching happens 
only at the root.  
The dual operator $\cost$ of $\st$, called {\bf branching corecurrence}, is defined in a symmetric way with the roles of the two players interchanged: here it is $\pp$ who can initiate new branches and for whom winning in one of the branches is sufficient. Alternatively, $\cost A$ can be defined as $\gneg \st\gneg A$. Again, winning $\cost A$ is easier for $\pp$ than winning the infinite disjunction $A\mld A\mld A\mld\ldots$ that we call {\bf parallel corecurrence} $\pcost A$ ($=\gneg\pst\gneg A$). To feel this, let us consider the bounded versions $\cost^2$ and $\pcost^2$ of $\cost$ and $\pcost$, in which the total number of allowed branches is limited to 2. 
We want to compare $\cost^2 B$ with $\pcost^2B$, i.e. with $B\mld B$, where\vspace{-3pt}  
\[B\ = (\chess\add\gneg\chess)\adc(\checkers\add\gneg\checkers).\vspace{-3pt}\]
Here is $\pp$'s strategy for $\cost^2 B$: Wait till $\oo$ chooses one of the $\adc$-conjuncts of $B$. 
Suppose the first conjunct is chosen (the other choice will be handled in a similar way). This brings the game down to $\chess\add\gneg\chess$. Now make a splitting move, thus creating two branches/copies of $\chess\add\gneg\chess$. In one copy choose $\chess$, and in the other copy choose $\gneg\chess$. From now on the game continues as a parallel play of $\chess$ and $\gneg\chess$, where it is sufficient for $\pp$ to win in one of the plays. Hence, applying the ``mimicking strategy" described in Section 1 for $\chess\mld\gneg\chess$ guarantees success. On the other hand, winning 
 $B\mld B$ is not easy. A possible scenario here is that $\oo$, by making different choices in the two 
disjuncts, brings the game down to $(\chess\add\gneg\chess)\mld (\checkers\add\gneg\checkers)$. Unless $\pp$ is a champion in either chess or checkers, (s)he may find it hard to win this game no matter what choices (s)he makes 
in its two disjuncts.

\section{Not-necessarily-constant games}

Classical logic identifies propositions with their truth values, so that there are exactly two propositions:
$\twg$ (true) and $\tlg$ (false), with the expressions ``snow is white" or ``2+2=4" simply being two different names
of the same proposition $\twg$, and ``elephants can fly" being one of the possible names of $\tlg$. Thinking of 
the classical propositions $\twg$ and $\tlg$ as the games $\twg$ and $\tlg$ defined in Section 2, classical propositions become a special --- elementary --- case of our constant games. It is not hard to see that our game operations 
$\gneg,\mlc,\mld,\mli$, when applied to $\twg$ and $\tlg$, again produce $\twg$ or $\tlg$, and exactly in the  
way their same-name classical counterparts do. Hence,
the  $\gneg,\mlc,\mld,\mli$-fragment of CL, restricted to elementary constant games, is nothing but classical propositional logic. The expressive power of propositional logic, however, is very limited. The more expressive version of classical logic --- first-order logic --- generalizes propositions to predicates. Let us fix two infinite sets of expressions: the set 
$\{v_1,v_2,\ldots\}$ of {\bf variables} and the set $\{1,2,,\ldots\}$ of {\bf constants}. Without loss of generality here we assume that the above collection of constants is exactly the universe of discourse
--- i.e. the set over which the variables range --- in all cases that we consider. By a {\bf valuation} we mean 
a function that sends each variable $x$ to a constant $e(x)$. In these terms, a classical {\bf predicate} $P$ can be understood as 
a function that sends each valuation $e$ to either $\twg$ (meaning that $P$ is true at $e$) or $\tlg$ (meaning that $P$ is false at $e$). Say, the predicate $x<y$ is the function that, for a valuation $e$, returns $\twg$ if  
$e(x)<e(y)$, and returns $\tlg$ otherwise. Propositions can then be thought of as special, {\em constant} cases of predicates --- predicates that return the same proposition for every valuation.

The concept of games that we define below generalizes constant games in exactly the same sense as the above classical concept of predicates generalizes propositions:

\begin{definition}\label{ngame}
A {\bf game} is a function from valuations to constant games. We write $e[A]$ (rather than $A(e)$) to denote the constant game returned by game $A$ for valuation $e$. Such a constant game $e[A]$ is said to be an {\bf instance} of $A$. 
\end{definition}

Just as this is the case with propositions versus predicates, constant games in the sense of Definition \ref{game} will
be thought of as special, {\em constant} cases of games in the sense of Definition \ref{ngame}. In particular, each constant game $A'$ is the game $A$ such that, for every valuation $e$,
$e[A]=A'$. From now on we will no longer distinguish between such $A$ and $A'$, so that, if $A$ is a constant game,
it is its own instance, with $A=e[A]$ for every $e$.      
 
We say that a game $A$ {\bf depends on} a variable $x$ iff there are two valuations $e_1,e_2$ that agree on all variables except $x$ such that $e_1[A]\not=e_2[A]$. Constant games thus do not depend on any variables. 

The notion of elementary game that we defined for constant games naturally generalizes to all games by stipulating that a given game is {\bf elementary} iff all of its instances are so. Hence, just as we identified classical propositions with constant elementary games, classical predicates from now on will be identified with elementary games. Say, $Even(x)$ is the elementary game such that $e[Even(x)]$ is the game $\twg$ if $e(x)$ is even, and the game $\tlg$ if $e(x)$ is odd. 

Any other concepts originally defined only for constant games can be similarly extended to all games. In particular, just as the propositional operations of classical logic naturally generalize to operations on predicates, so do our game operations
from Section 2. This is done by simply stipulating that $e[\ldots]$ commutes with all of those operations: $\gneg A$ is 
the game such that, for every $e$, $e[\gneg A]=\gneg e[A]$; $A\adc B$ is the game such that,
for every $e$, $e[A\adc B]=e[A]\adc e[B]$; etc. A little caution is necessary when generalizing the operation of prefixation this way. As we remember, for a constant game $A$, $\seq{\Phi}A$ is defined only when $\Phi$ is a legal 
position of $A$. So, for $\seq{\Phi}A$ to be defined for a not-necessarily-constant game $A$, $\Phi$ should be 
a legal position of every instance of $A$. 
Once this condition is satisfied, $\seq{\Phi}A$ is defined as the game such that, for every valuation $e$, $e[\seq{\Phi}A]=\seq{\Phi}e[A]$. 

To generalize the standard operation of substitution of variables to games, let us agree that by a {\bf term} we mean either 
a variable or a constant; the domain of each valuation $e$ is extended to all terms by stipulating that, for any constant $c$, $e(c)=c$. 

\begin{definition}\label{sov}
Let $A$ be a game, $x_1,\ldots,x_n$ pairwise distinct variables, and $t_1,\ldots,t_n$ any (not necessarily distinct) terms.
The result of {\bf substituting $x_1,\ldots,x_n$ by $t_1,\ldots,t_n$ in $A$}, denoted $A(x_1/t_1,\ldots,x_n/t_n)$, is defined by stipulating that, for every valuation $e$, $e[A(x_1/t_1,\ldots,x_n/t_n)]=e'[A]$, where $e'$ is the valuation for which we have:\vspace{3pt}

$\begin{array}{l}
1.\ \ e'(x_1)=e(t_1),\ \ldots,\ e'(x_n)=e(t_n);\\
2.\  \ \mbox{for every variable $y\not\in\{x_1,\ldots,x_n\}$, $e'(y)=e(y)$.} 
\end{array}$
\end{definition}

Intuitively $A(x_1/t_1,\ldots,x_n/t_n)$ is $A$ with $x_1,\ldots,x_n$ remapped to $t_1,\ldots,t_n$, respectively. 
Say, if $A$ is the predicate/elementary game $x<y$, then $A(x/y,y/x)$ is $y<x$, $A(x/y)$ is $y<y$, $A(y/3)$ is $x<3$, and $A(z/3)$ --- where $z$ is different from $x,y$ --- remains $x<y$ because $A$ does not depend on $z$.

Following the standard readability-improving practice established in the literature for predicates, we will often fix a tuple $(x_1,\ldots,x_n)$ of pairwise distinct variables for a game $A$ and write $A$ as $A(x_1,\ldots,x_n)$. 
It should be noted that when doing so, by no means do we imply that $x_1,\ldots,x_n$ are of all of 
(or only) the variables on which $A$ depends. Representing $A$ in the form $A(x_1,\ldots,x_n)$ sets a context in which we can write $A(t_1,\ldots,t_n)$ to mean the same as the more clumsy expression $A(x_1/t_1,\ldots,x_n/t_n)$. So, if the game $x<y$ is represented as $A(x)$, then $A(3)$ will mean $3<y$ and $A(y)$ mean 
$y<y$. And if the same game is represented as $A(y,z)$ (where $z\not=x,y$), then $A(z,3)$ means $x<z$ while $A(y,3)$ again means $x<y$.

The entities that in common language we call ``games" are at least as often non-constant as constant. Chess  
is a classical example of a constant game. On the other hand, many of the card games --- including solitaire games where only 
one player is active --- are more naturally represented as non-constant games: each session/instance of such a game is set 
by a particular permutation of the card deck, and thus the game can be understood as a game that depends on a variable $x$
 ranging over the possible settings of the deck or certain portions of it. Even the game of checkers --- another ``classical example" of a constant game --- has a natural non-constant generalization 
$\checkers(x)$ (with $x$ ranging over positive even integers), meaning a play on the board of size $x\times x$ where,
in the initial position,  
the first $\frac{3}{2}x$ black cells are filled with white pieces and the last $\frac{3}{2}x$ black cells with black pieces.  Then the 
ordinary checkers can be written as $\checkers(8)$. Furthermore, the numbers of 
pieces of either color also can be made variable, getting an even more general game $\checkers(x,y,z)$, with the ordinary checkers being 
the instance $\checkers(8,12,12)$ of it. By allowing rectangular- (rather than just square-) shape boards, we would get a game that depends on four variables, etc. Computability theory texts also often appeal to non-constant games to illustrate 
certain complexity-theory concepts such as alternating computation or PSPACE-completeness. The {\em Formula Game} or 
{\em Generalized Geography} (\cite{Sip77}, Section 8.3) are typical examples. Both can be understood as games that depend on a variable $x$, with $x$ ranging over quantified Boolean formulas in Formula Game and over directed graphs in Generalized Geography.

A game $A$ is said to be {\bf unistructural in} a variable $x$ iff, for every two valuations $e_1$ and $e_2$ that agree on all variables except $x$, we have $\legal{e_1[A]}{}=\legal{e_2[A]}{}$. And $A$ is (simply) {\bf unistructural} iff 
$\legal{e_1[A]}{}=\legal{e_2[A]}{}$ for any two valuations $e_1$ and $e_2$. Intuitively, a unistructural game is a game whose every instance has the same structure (the $\legal{}{}$ component). And $A$ is unistructural in $x$ iff the structure 
of an instance $e[A]$ of $A$ does not depend on how $e$ evaluates the variable $x$. Of course, every constant or elementary game is unistructural, and every unistructural game is unistructural in all variables. The class of unistructural games can be shown to be closed under all of our game operations (Theorem \ref{predstat}). While natural examples of non-unistructural games exist such as the games mentioned in the above paragraph, virtually all of the other examples of particular games discussed elsewhere in the present paper are unistructural. In fact, every non-unistructural game can be 
rather easily rewritten into an equivalent (in a certain reasonable sense) unistructural game. One of the standard ways to 
convert a non-unistructural game $A$ into a corresponding unistructural game $A'$ is to take the  
union  (or anything bigger) $U$ of the structures of all instances of $A$ to be the common-for-all-instances structure of $A'$, and then extend the $\win{}{}$ function of each instance $e[A]$ of $A$ to $U$ by stipulating that, if $\Gamma\not\in \legal{e[A]}{}$, then the player who made the first illegal (in the sense of $e[A]$) move is the loser in $e[A']$. So, say, in the unistructural version of generalized checkers, an attempt by a player to move to or from a non-existing cell would result in a loss for that player but otherwise considered a legal move. In view of these remarks, if the reader feels more comfortable this way, without much loss of generality (s)he can always understand ``game" as ``unistructural game".

Now we are ready to define quantifier-style operations on games. 
The blind group $\cla x,\cle x$ of quantifiers is only defined  for games that are unistructural in $x$.

\begin{definition}\label{opq}
Below $A(x)$ is an arbitrary game that in Clauses 5 and 6 is assumed to be unistructural in $x$. $e$ ranges over all valuations. Just as in Definition \ref{op}, $\Phi$ ranges over nonempty positions, and $\Gamma$ ranges over the legal runs of the game that is being defined.
The notation $\Phi^{c.}$ in clauses 3 and 4 means the result of removing from $\Phi$ all labeled moves 
except those of the form $\xx c.\alpha$ ($\xx\in\{\pp,\oo\}$), and then deleting the prefix `$c.$' in the remaining moves, i.e. replacing each such $\xx c.\alpha$ by $\xx\alpha$. Similarly for $\Gamma^{c.}$.

\begin{enumerate}
\item {\bf Choice universal quantification} $\ada xA(x)$:
\begin{itemize}
\item $\Phi\in\legal{e[\adai xA(x)]}{}$ iff $\Phi=\seq{\oo c,\Psi}$, where $c$ is a constant and 
$\Psi\in\legal{e[A(c)]}{}$; 
\item $\win{e[\adai xA(x)]}{}\seq{\Gamma}=\oo$ iff $\Gamma=\seq{\oo c,\Upsilon}$, where $c$ is a constant and $\win{e[A(c)]}{}\seq{\Upsilon}=\oo$.
\end{itemize}

\item {\bf Choice existential quantification} $\ade xA(x)$:
\begin{itemize}
\item $\Phi\in\legal{e[\adei xA(x)]}{}$ iff $\Phi=\seq{\pp c,\Psi}$, where $c$ is a constant and 
$\Psi\in\legal{e[A(c)]}{}$; 
\item $\win{e[\adei xA(x)]}{}\seq{\Gamma}=\pp$ iff $\Gamma=\seq{\pp c,\Upsilon}$, where $c$ is a constant and $\win{e[A(c)]}{}\seq{\Upsilon}=\pp$.
\end{itemize}

\item {\bf Parallel universal quantification} $\mla xA(x)$:
\begin{itemize}
\item $\Phi\in \legal{e[\mlai xA(x)]}{}$ iff every move of $\Phi$ starts with `$c.$' for some constant $c$  and,
for each such $c$,  $\Phi^{c.}\in\legal{e[A(c)]}{}$; 
\item $\win{e[\mlai xA(x)]}{}\seq{\Gamma}=\pp$ iff, for each constant $c$, $\win{e[A(c)]}{}\seq{\Gamma^{c.}}=\pp$.
\end{itemize}

\item {\bf Parallel existential quantification} $\mle xA(x)$:
\begin{itemize}
\item $\Phi\in \legal{e[\mlei xA(x)]}{}$ iff every move of $\Phi$ starts with `$c.$' for some constant $c$  and,
for each such $c$,  $\Phi^{c.}\in\legal{e[A(c)]}{}$; 
\item $\win{e[\mlei xA(x)]}{}\seq{\Gamma}=\oo$ iff, for each constant $c$, $\win{e[A(c)]}{}\seq{\Gamma^{c.}}=\oo$.
\end{itemize}

\item {\bf Blind universal quantification} $\cla xA(x)$:
\begin{itemize}
\item $\Phi\in \legal{e[\clai xA(x)]}{}$ iff $\Phi\in \legal{e[A(x)]}{}$; 
\item $\win{e[\clai xA(x)]}{}\seq{\Gamma}=\pp$ iff, for each constant $c$, $\win{e[A(c)]}{}\seq{\Gamma}=\pp$.
\end{itemize}

\item {\bf Blind existential quantification} $\cle xA(x)$:
\begin{itemize}
\item $\Phi\in \legal{e[\clei xA(x)]}{}$ iff $\Phi\in \legal{e[A(x)]}{}$;
\item $\win{e[\clei xA(x)]}{}\seq{\Gamma}=\oo$ iff, for each constant $c$, $\win{e[A(c)]}{}\seq{\Gamma}=\oo$.
\end{itemize}
\end{enumerate}
\end{definition}

Thus, $\ada xA(x)$ and $\mla xA(x)$ are nothing but  $A(1)\adc A(2)\adc\ldots$ and  $A(1)\mlc A(2)\mlc\ldots$, respectively. Similarly, $\ade$ and $\mle$ are ``big brothers" of $\add$ and $\mld$.
As for $\cla xA(x)$, as explained in Section 1, winning it for $\pp$ (resp. $\oo$) means winning $A(x)$, at once, for all (resp. some)  
possible values of $x$ without knowing the actual value of $x$. Playing or evaluating a game generally  might be 
impossible or meaningless without knowing what moves are available/legal. Therefore our definition of $\cla xA(x)$
and $\cle xA(x)$ insists that the move legality question should not depend on the (unknown) value of $x$, i.e. that $A(x)$ should be unistructural in $x$. 


As we did in Exercise \ref{ex01}, one can easily verify the following interdefinabilities:
\[\begin{array}{ll}
\ade xA(x)=\gneg\ada x\gneg A(x); & \ada xA(x)=\gneg\ade x\gneg A(x);\\
\mle xA(x)=\gneg\mla x\gneg A(x); & \mla xA(x)=\gneg\mle x\gneg A(x);\\
\cle xA(x)=\gneg\cla x\gneg A(x); & \cla xA(x)=\gneg\cle x\gneg A(x).
\end{array}\]
 
\begin{exercise}\label{quantex}
Let $\mbox{\em Odd}(x)$ be the predicate ``$x$ is odd". Verify that:

1. $\seq{\oo 3,\pp 1}$ is a legal run of $\ada x\bigl(\mbox{\em Odd}(x)\add \gneg \mbox{\em Odd}(x)\bigr)$ won by $\pp$.

2. $\cla x\bigl(\mbox{\em Odd}(x)\add \gneg \mbox{\em Odd}(x)\bigr)$ has exactly three legal runs: $\emptyrun$, $\seq{\pp 1}$ and $\seq{\pp 2}$, all lost by $\pp$. $\cle x\bigl(\mbox{\em Odd}(x)\add \gneg \mbox{\em Odd}(x)\bigr)$ has the same legal runs, with $\emptyrun$ won by $\oo$ and the other two by $\pp$. 

3. $\seq{\pp 9.1}$ is a legal run of $\mle x\bigl(\mbox{\em Odd}(x)\add \gneg \mbox{\em Odd}(x)\bigr)$ won by $\pp$.

4. $\seq{\pp 1.1,\pp 2.2,\pp 3.1,\pp 4.2,\pp 5.1,\pp 6.2,\ldots}$ is a legal run of $\mla x\bigl(\mbox{\em Odd}(x)\add \gneg \mbox{\em Odd}(x)\bigr)$
won by $\pp$. On the other hand,  every finite initial segment of this infinite run is lost by $\pp$. 
\end{exercise}

\begin{exercise} Verify that, for every game $A(x)$, we have: 

1. Where $c$ is an arbitrary constant, \ $\seq{\oo c}\ada xA(x)= A(c)$ and $\seq{\pp c}\ade xA(x)= A(c)$.

2. Where $A(x)$ is unistructural in $x$ and $\Phi$ is a legal position of all instances of $A(x)$, 
$\seq{\Phi}\cla xA(x)=\cla x\seq{\Phi}A(x)$ and  $\seq{\Phi}\cle xA(x)=\cle x\seq{\Phi}A(x)$.
\end{exercise}

The results of the above exercise will help us visualize legal runs of $\cla,\cle,\ada,\ade$-combinations of games in the style of the earlier Example \ref{nov26}:

\begin{example} Let $E(x,y)$ be the predicate ``$x+y$ is even", and $G_0$ be the game 
$\cla x\bigl(\bigl(E(x,4)\add\gneg E(x,4)\bigr)\mli\ada y\bigl(E(x,y)\add\gneg E(x,y)\bigr)\bigr)$, i.e.  
$\cla x\bigl(\bigl(\gneg E(x,4)\adc E(x,4)\bigr)\mld\ada y\bigl(E(x,y)\add\gneg E(x,y)\bigr)\bigr)$. Then 
$\seq{\oo 2.7,\oo 1.2,\pp 2.1}$ is a legal run of $G_0$, to which corresponds the following sequence of games:\vspace{3pt}

$\begin{array}{lll}
G_0: & \cla x\bigl(\bigl(\gneg E(x,4)\adc E(x,4)\bigr)\mld\ada y\bigl(E(x,y)\add\gneg E(x,y)\bigr)\bigr),\ \  &  
\mbox{}\\
G_1: & \cla x\bigl(\bigl(\gneg E(x,4)\adc E(x,4)\bigr)\mld\bigl(E(x,7)\add\gneg E(x,7)\bigr)\bigr), &  
\mbox{i.e. $\seq{\oo 2.7}G_0$;}\\
G_2: & \cla x\bigl(E(x,4)\mld\bigl(E(x,7)\add\gneg E(x,7)\bigr)\bigr), &  
\mbox{i.e. $\seq{\oo 1.2}G_1$;}\\
G_3: & \cla x\bigl(E(x,4)\mld E(x,7)\bigr), & 
\mbox{i.e. $\seq{\pp 2.1}G_2$.}\vspace{2pt} 
\end{array}$

The run hits the true proposition $\cla x\bigl(E(x,4)\mld E(x,7)\bigr)$ and hence is won by $\pp$. Note that ---
  just as this is the case with all non-choice operations ---  the $\cla,\cle$-structure of a game persists throughout a run.
\end{example}

When visualizing $\mla,\mle$-games in a similar style, we are better off representing them as infinite conjunctions/disjunctions. Of course, putting infinitely many conjuncts/disjuncts on paper would be no fun. But, luckily,
in every position of such (sub)games $\mla xA(x)$ or $\mle xA(x)$ only a finite number of conjuncts/disjuncts would be ``activated", i.e. have a non-$A(c)$ form, so that all of the other, uniform, conjuncts can be combined into blocks and represented, say, through an ellipsis, or through expressions such as $\mla m\leq x\leq n A(x)$ or $\mla x\geq mA(x)$. 
 Once $\mla,\mle$-formulas are represented as parallel conjunctions/disjunctions, 
we can apply the results of Exercise \ref{mar12}(3) --- now generalized to infinite conjunctions/disjunctions --- to visualize runs. For example, the legal run $\seq{\pp 9.1}$ of game $H_0=\mle x\bigl(\mbox{\em Odd}(x)\add \gneg \mbox{\em Odd}(x)\bigr)$ from Exercise \ref{quantex}(3) will be represented as follows:\vspace{3pt}

$\begin{array}{lll}
H_0: & \mle x\geq 1\bigl(\mbox{\em Odd}(x)\add \gneg \mbox{\em Odd}(x)\bigr); &  \\
H_1: & \mle 1\leq x\leq 8\bigl(\mbox{\em Odd}(x)\add \gneg \mbox{\em Odd}(x)\bigr)\mld \mbox{\em Odd}(9)\mld 
\mle x\geq 10\bigl(\mbox{\em Odd}(x)\add \gneg \mbox{\em Odd}(x)\bigr), & \mbox{i.e. $\seq{\pp 9.1}H_0$.}\vspace{2pt}
\end{array}$

\noindent And the infinite legal run $\seq{\pp 1.1,\pp 2.2,\pp 3.1,\pp 4.2,\pp 5.1,\pp 6.2,\ldots}$ of game 
$J_0=\mla x\bigl(\mbox{\em Odd}(x)\add \gneg \mbox{\em Odd}(x)\bigr)$ from Exercise \ref{quantex}(4) will be represented as follows:\vspace{3pt}

$\begin{array}{lll}
J_0: & \mla x\geq 1\bigl(\mbox{\em Odd}(x)\add \gneg \mbox{\em Odd}(x)\bigr); & \\
J_1: & \mbox{\em Odd}(1) \mlc\mla x\geq 2\bigl(\mbox{\em Odd}(x)\add \gneg \mbox{\em Odd}(x)\bigr), & 
\mbox{i.e. $\seq{\pp 1.1}J_0$;}\\
J_2: & \mbox{\em Odd}(1) \mlc \gneg\mbox{\em Odd}(2)\mlc \mla x\geq 3\bigl(\mbox{\em Odd}(x)\add \gneg \mbox{\em Odd}(x)\bigr), & \mbox{i.e. $\seq{\pp 2.2}J_1$;}\\
J_3: & \mbox{\em Odd}(1) \mlc \gneg \mbox{\em Odd}(2)\mlc \mbox{\em Odd}(3)\mlc\mla x\geq 4\bigl(\mbox{\em Odd}(x)\add \gneg \mbox{\em Odd}(x)\bigr), \ \ & \mbox{i.e. $\seq{\pp 3.1}J_2$;}\\
& \mbox{...etc.} &
\end{array}$
  
\section{Interactive computational problems}\label{icp}

Various sorts of games have been extensively studied in both logical and theoretical computer science literatures. 
The closest to our present approach to games appears to be Blass's \cite{Bla92} model, and less so the models 
proposed later within the `game semantics for linear logic' line by Abramsky, Jagadeesan, Hyland, Ong and others. 
See Section 27 of \cite{Jap03} for a discussion of how other game models compare with our own, and what the 
crucial advantages of our approach to games are that turn the corresponding logic into a logic of computability --- something that is no longer ``just a game". One of the main distinguishing features of our  games 
is the absence of what in \cite{Ben01} is called {\em procedural rules} --- rules strictly regulating who and when should move, the most standard procedural rule being the one according to which the players should take turns in alternating order. 
In our games, either player is free to make any (legal) move at any time. Such games can be called {\bf free}, while games where in any given situation only one of the players is allowed to move 
called {\bf strict}. Strict games can be thought of as special cases of our free games, where the structure ($\legal{}{}$) component is such that in any given position at most one of the players has legal moves.
Our games are thus most general of all two-player, two-outcome games. This makes them the most powerful and flexible modeling tool for interactive tasks.
It also makes our definitions of game operations as simple, compact and natural as they could be, and allows us to adequately capture certain intended intuitions associated with those operations. Consider the game {\em Chess}$\mlc${\em Chess}. Assume an agent plays this two-board game over the Internet against two independent adversaries that, together, form the (one) environment for the agent. Playing white on both boards, in the initial position of this game only the agent has legal moves. But once such a move is made, say, on the left board, the picture changes. Now both the agent and the environment have legal moves: the agent may make another opening move on the right board, while the environment --- in particular, adversary \#1 --- may make a reply move on the left board. This is a situation where which player `can move' is no longer strictly determined, so the next player to move will be the one who can or wants to act sooner. A strict-game approach would impose some additional conditions uniquely determining the next player to move. Such conditions would most likely be artificial and not quite adequate, for the situation we are trying to model is a concurrent play on two boards against two independent adversaries, and we cannot or should not expect any coordination between their actions. Most of the compound tasks that we perform in everyday life are free rather than strict, and so are most computer communication/interaction protocols. A strict understanding of $\mlc$ would essentially mean some sort of an (in a sense interlaced but still) sequential rather than truly parallel/concurrent combination of tasks, where no steps in one component can be made until receiving a response in the other component, contrary to the very (utility-oriented) idea of parallel/distributed computation.   

Our class of free games is obviously general enough to model anything that we would call a (two-agent, 
two-outcome) interactive problem.  However, it is too general. There are games where the chances of a player to succeed essentially depend on the relative speed at which its adversary responds, and
we do not want to consider that sort of games meaningful computational problems. 
A simple example would be a game where all moves are legal and that is won by the player who moves first. This is merely a contest of speed.  
Below we define a subclass of games called static games. Intuitively, they are games where speed is irrelevant: in order to succeed (play legal and win), only matters {\em what} to do (strategy) rather than {\em how fast} to do (speed). In particular, if a player can succeed when acting fast in such a game, it will remain equally successful acting the same way but slowly. This releases the player from any pressure for time and allows it to select its own pace for the game. 

We say that a run $\Upsilon$ is a {\bf $\xx$-delay} of a run $\Gamma$ iff the following two conditions are satisfied:\vspace{-5pt}
\begin{itemize}
\item for each player $\xx'$, the subsequence of $\xx'$-labeled moves of $\Upsilon$ is the same as that of $\Gamma$, and\vspace{-5pt}
\item for any $n,k\geq 1$, if the $n$th $\xx$-labeled move is made later than (is to the right of) the $k$th non-$\xx$-labeled move in $\Gamma$, then so is it in $\Upsilon$.\vspace{-5pt}
\end{itemize}
\noindent This means that in  $\Upsilon$  each player has made the same sequence of moves as in $\Gamma$, only, in $\Upsilon$, $\xx$ might have been acting with some delay.
Then we say that a constant game  $A$ is {\bf static}\label{static} iff, whenever a run $\Upsilon$ is a $\xx$-delay of 
a run $\Gamma$, we have:\vspace{-5pt}
\begin{itemize}
\item if $\Gamma$ is a $\xx$-legal run of $A$, then so is $\Upsilon$,\footnote{This first condition was a derivable one in the presentation chosen in \cite{Jap03}. See the footnote on page \pageref{ftn}.}  and\vspace{-5pt}\label{stg}
\item if $\Gamma$ is a $\xx$-won run of $A$, then so is $\Upsilon$.\vspace{-5pt}
\end{itemize}
This definition extends to all games by stipulating that a (not-necessarily-constant) game is {\bf static} iff all of its instances are so.

Now we are ready to formally clarify what we mean by interactive computational problems: an {\bf interactive computational problem (ICP)} is a static game, and from now on we will be using the terms ``ICP" (or simply ``problem") and ``static game" 
interchangeably. 
This terminology is justified by one of the two main theses on which CL relies philosophically: {\em the concept of static games is an adequate 
formal counterpart of our intuitive notion of ``pure", speed-independent interactive computational problems}. 
See Section 4 of \cite{Jap03} for a detailed discussion and examples in support of this thesis. According to the second thesis, 
{\em the concept of computability/winnability of static games, defined in the next section, is an adequate formal counterpart of our intuitive notion of effective solvability of speed-independent interactive problems}. This is thus an 
interactive version of the Church-Turing thesis. 
 
\begin{theorem}\label{predstat} \ 

1. Every elementary game is static and unistructural.

2. All of our game operations --- $\gneg$, $\mlc$, $\mld$, $\mli$, $\adc$, $\add$, $\st$, $\cost$, $\pst$, $\pcost$, $\ada$, $\ade$, $\cla$, $\cle$, $\mla$, $\mle$, 
prefixation and substitution of variables --- preserve both the static and the unistructural properties of games.
\end{theorem}

The first clause of this theorem is straightforward; the second clause has been proven in \cite{Jap03} (Theorem 14.1) for all operations except $\pst,\pcost,\mla$ and $\mle$ that were not officially introduced there but that 
can be handled in exactly the same way as $\mlc,\mld$. 

In view of Theorem \ref{predstat}, the closure of the set of all predicates under all of our game operations forms a natural class $\cal C$ of unistructural ICPs. For a reader who has difficulty in comprehending the concept of static games, it is perfectly safe to simply think of ICPs as elements of $\cal C$: even though the class $\cal ICP$ of all ICPs is essentially wider than $\cal C$, virtually all of our results --- in particular, completeness results --- remain valid with $\cal ICP$ 
restricted to $\cal C$.  
Class $\cal C$ has a number of nice features. 
Among them, together with unistructurality, is the effectiveness of the structure of any $A\in{\cal C}$, in the sense that 
the question whether a given move is legal in a given position is decidable --- in fact, decidable rather efficiently. 
 
\section{Interactive computability}\label{cp}
Now that we know what ICPs are, it is time to clarify what their computability means. The definitions given in this section are semiformal. All of the omitted technical details are rather standard or irrelevant and can be easily restored by anyone familiar with Turing machines. 
If necessary, the corresponding detailed definitions can be found in Part II of \cite{Jap03}. 
  
As we remember, the central point of our philosophy is to require that agent $\pp$ be implementable as a computer program, with effective and fully determined behavior. On the other hand, the behavior of agent $\oo$ can be arbitrary. This intuition is captured by the model of interactive computation where $\pp$ is formalized as what we call 
{\bf HPM}.\footnote{HPM stands for `Hard-Play Machine'. See \cite{Jap03} for a (little long) story about why ``hard".} 

An HPM $\cal M$ is a Turing machine that, together with an ordinary read/write {\em work tape}, has two additional, read-only tapes: the {\em valuation tape} and the {\em run tape}. The presence of these two tapes is related to the fact that 
the outcome of a play over a given game depends on two parameters: (1) valuation and (2) the run that is generated 
in the play. $\cal M$ should have full access to information about these two parameters, and this information is provided 
by the valuation and run tapes: the former spells a (the ``actual") valuation $e$ by listing constants in the lexicographic order of the corresponding variables, and the latter spells, at any given time, the current position, i.e. the sequence of the (labeled) moves made by the two players so far, in the order in which those moves have been made. Thus, both of these two tapes can be considered input tapes. The reason for our choice to keep them separate is the difference in the nature of the input that they provide. Valuation is a {\em static} input, known at the very beginning of a computation/play and remaining unchanged throughout the subsequent process. On the other hand, the input provided by the run tape is {\em dynamic}: 
every time one of the players makes a move, the move (with the corresponding label) is appended to the content of this  
tape, with such content being unknown and hence blank at the beginning of interaction. Technically the run tape is read-only: the machine has unlimited read access to this (as well as to the valuation) tape, but it cannot write directly on it. 
Rather, $\cal M$ makes a move $\alpha$ by constructing it at the beginning of its work tape, delimiting its end with a 
blank symbol, and entering one of the specially designated states called {\em move states}. Once this happens, $\pp\alpha$ is automatically appended to  the current  position spelled on the run tape. 
While the frequency at which the machine can make moves is naturally limited by its clock cycle time (the time each computation step takes), there are no limitations to how often the environment can make a move, so, during one computation step of the machine, any finite number of any moves by the environment can be appended to 
the content of the run tape. This corresponds to the intuition that not only the strategy, but also the relative speed of the  environment can be arbitrary. For technical clarity, we can assume that the run tape remains stable during a clock cycle and is updated only on a transition from one cycle to another, on which event the moves (if any) by the two players appear on it at once in the order that they have been made. As we may guess, the computing power of the machine is rather rigid with respect to how this sort of technical details are arranged, and such details can be safely suppressed.

A {\bf configuration}\label{y10} of $\cal M$ is defined in the standard way: this is a full description of the (``current") state of the machine, the locations of its three scanning heads
and the contents of its tapes, with the exception that, in order to make finite descriptions of configurations possible, we do not formally include a description of the unchanging 
(and possibly essentially infinite) content of the valuation tape as a part of configuration, but rather account for it in our definition of computation branch as this will be seen below. 
The {\em initial configuration} is the configuration where $\cal M$ is in its start state and the work and run tapes are empty. A configuration $C'$ is said to be an {\bf $e$-successor} of configuration $C$ if, when valuation $e$ is spelled on the valuation tape,  $C'$ can legally follow $C$ in the standard  sense, based on the transition function (which we assume to be deterministic) of the machine and accounting for the possibility of the above-described nondeterministic updates of the content of the run tape. An {\bf $e$-computation branch}\label{y4} of $\cal M$ is a sequence of configurations of $\cal M$ where the first
configuration is the   initial configuration and every other configuration is an $e$-successor of the previous one.
Thus, the set of all $e$-computation branches captures all possible scenarios (on valuation $e$) corresponding to different behaviors by $\oo$.
Each $e$-computation branch $B$ of $\cal M$ incrementally spells --- in the obvious sense --- a run $\Gamma$ on the run tape, which we call the {\bf run spelled by $B$}.\label{x53} 

\begin{definition}\label{feb2}
 For ICPs $A$ and $B$ we  say that:

1. An HPM $\cal M$ {\bf computes} ({\bf solves}, {\bf wins}) $A$ iff, for every valuation $e$, whenever 
$\Gamma$ is the run spelled by some $e$-computation branch of $\cal M$, $\Gamma$ is a $\pp$-won legal run of $e[A]$ as long as it is $\oo$-legal. 

2. $A$ is {\bf computable} iff there is an HPM that computes $A$. Such an HPM is said to be a {\bf solution} to $A$.

\mbox{3. $A$} is \ {\bf reducible}  to $B$ iff $B\mli A$ is computable.  An HPM that computes $B\mli A$ is said to be a {\bf reduction} of $A$ to $B$. 

4. $A$ and $B$ are {\bf equivalent} iff $A$ is reducible to $B$ and $B$ is reducible to $A$. 
\end{definition}

One of the most appealing known models of interactive computation is {\em Persistent Turing Machines} (Goldin \cite{Gol00}). PTMs are defined as Turing machines where the content of the work tape persists between an output and the subsequent input events (while an ordinary Turing machine cleans up the tape and starts from scratch on every new input). The
PTM model appears to be optimal for what is called {\em sequential interactive computations} 
\cite{Gol?}, which into our terms translate as plays over games with strictly alternating legal moves by the two players,
always by the environment to start. Our HPM model sacrifices some of the niceties of PTMs in its ambition to capture the wider class of free games and, correspondingly, not-necessarily-sequential interactive computations.  

Just as the Turing machine model, our HPM model, as this was noted, is highly rigid with respect to reasonable technical variations. Say, a model where only environment's moves are visible to the machine  
yields the same class of computable ICPs. Similarly, there is no difference between whether we allow the scanning heads on the valuation and run tapes to move in either or only one (left to right) direction. Another variation is the one where an attempt by either 
player to make an illegal move has no effect: such moves are automatically rejected and/or filtered out by some interface hardware or software and thus illegal runs are never generated. Obviously in such a case a minimum requirement would be that 
the question of legality of moves be decidable. This again yields a model equivalent to HPM.

\section{The propositional logic of computability}\label{sul}

Among the main technical goals of CL at its present stage of development is to axiomatize the set of valid principles of
computability or various natural fragments of that set. This is a challenging but promising task. Some positive results
in this direction have already been obtained, yet more results are still to come. We start our brief survey of known results  at the simplest, propositional level. The system axiomatizing the most basic fragment of propositional computability 
logic is called $\propel$. Its language extends that of classical propositional logic by incorporating into it two additional connectives:
$\adc$ and $\add$. As always, there are infinitely many {\bf atoms} in the language, for which we will be using the letters
$p,q,r,\ldots$ as metavariables. Atoms are meant to represent elementary games. The two  atoms: $\twg$ and $\tlg$ have a special status in that their interpretation is fixed. Therefore we call them {\bf logical} to distinguish them from all other atoms that are called {\bf nonlogical}. Formulas of this language, referred to as {\bf $\propel$-formulas}, are built from atoms in the standard way: the class of $\propel$-formulas 
is defined as the smallest set of expressions such that all atoms are in it and, if $F$ and $G$ are in it, then so are 
$\gneg (F)$, $(F)\mlc (G)$, $(F)\mld (G)$, $(F)\mli(G)$, $(F)\adc(G)$, $(F)\add(G)$. For better readability, we will often omit some parentheses in formulas by standard conventions.

An {\bf interpretation}, corresponding to what is more often called ``{\em model}\hspace{1pt}" in classical logic, is a function $^*$ that 
sends every nonlogical atom $p$ to an elementary game $p^*$. The mapping $^*$ uniquely extends to all $\propel$-formulas by stipulating that it commutes with all connectives, i.e. respects their meaning as game operations. That is, we have: 
$\twg^*=\twg$; $\tlg^*=\tlg$; $(\gneg G)^*=\gneg (G^*)$; $(G\mlc H)^*=(G^*)\mlc (H^*)$; $(G\mld H)^*=(G^*)\mld (H^*)$; $(G\mli H)^*=(G^*)\mli (H^*)$; $(G\adc H)^*=(G^*)\adc (H^*)$; $(G\add H)^*=(G^*)\add (H^*)$.

When $F^*=A$, we say that $^*$ {\bf interprets} $F$ as $A$. 
We say that a $\propel$-formula $F$ is {\bf valid} iff, for every interpretation $^*$, the ICP $F^*$ is computable. 
Thus, valid $\propel$-formulas are exactly the ones representing ``always-computable" problems, i.e. ``valid principles of computability".
 
Note that, despite the 
fact that we refer to $\propel$ as a ``propositional logic", interpretations of its formulas go beyond constant games, let alone propositions. This is so because our definition of interpretation does not insist that atoms be interpreted 
as constant games. Rather, for the sake of generality, it lets them represent any predicates.

To axiomatize the set of valid $\propel$-formulas, we need some preliminary terminology. Understanding $F\mli G$ as an abbreviation of $(\gneg F)\mld G$,  by a {\bf positive} (resp. {\bf negative)} {\bf occurrence} of a subexpression we mean an occurrence that is in the scope of an even (resp. odd) number of occurrences of $\gneg$. In the context of the language of $\propel$, 
by an {\bf elementary formula} we mean a formula not containing choice operators $\adc$, $\add$, i.e. a formula of classical propositional logic. A {\bf surface occurrence} of a subexpression means an occurrence that is not in the scope of choice operators. The {\bf elementarization} of a $\propel$-formula $F$ is the result of replacing in $F$ every surface occurrence  of the form $G\adc H$ by $\twg$ and every surface occurrence of the form $G\add H$ by $\tlg$.
A $\propel$-formula is said to be {\bf stable}\label{x57} iff its elementarization is a valid formula (tautology) of classical logic. Otherwise it is {\bf instable}.\label{x32}

With ${\cal P}\mapsto C$ here and later meaning ``from premise(s) ${\cal P}$ conclude $C$", deductively {\bf $\propel$} is given by the following two rules of inference:  
\begin{description}
\item[Rule (a):]  $\vec{H}\mapsto F$, where $F$ is stable and $\vec{H}$ is a set of formulas such that, 
whenever $F$ has a positive (resp. negative)  surface occurrence of a subformula $G_1\adc G_2$ (resp. $G_1\add G_2$), for each 
$i\in\{1,2\}$, $\vec{H}$ contains the result of replacing that occurrence in $F$ by $G_i$.
\item[Rule (b):]  $H\mapsto F$, where $H$ is the result of replacing in $F$ a negative (resp. positive) surface occurrence of a subformula $G_1\adc G_2$ (resp. $G_1\add G_2$) by $G_i$ for some $i\in\{1,2\}$.
\end{description}

Axioms are not explicitly stated, but note that the set $\vec{H}$ of premises of Rule {\bf (a)} can be empty, in which case the conclusion $F$ of that rule acts as an axiom.  A rather unusual logic, isn't it? Let us play with it a little to get a syntactic feel of it. Below, $p,q,r$ are pairwise distinct nonlogical atoms.

\begin{example}\label{ex03} The following is a $\propel$-proof of $\bigl((p\mli q)\adc (p\mli r)\bigr)\mli\bigr(p\mli (q\adc r)\bigr)$:

$\begin{array}{ll}
\mbox{1. $(p\mli q)\mli(p\mli q)$} & \mbox{(from $\{\}$ by Rule {\bf (a)})}\\
\mbox{2. $\bigl((p\mli q)\adc (p\mli r)\bigr)\mli\bigr(p\mli q\bigr)$} & \mbox{(from 1 by Rule {\bf (b)})}\\
\mbox{3. $(p\mli r)\mli(p\mli r)$} & \mbox{(from $\{\}$ by Rule {\bf (a)})}\\ 
\mbox{4. $\bigl((p\mli q)\adc (p\mli r)\bigr)\mli\bigr(p\mli r\bigr)$} & \mbox{(from  3 by Rule {\bf (b)})}\\
\mbox{5. $\bigl((p\mli q)\adc (p\mli r)\bigr)\mli\bigr(p\mli (q\adc r)\bigr)$} & \mbox{(from \{2,4\} by Rule {\bf (a)})}
\end{array}$\vspace{3pt}

 On the other hand, $\propel$ does not prove    
$\bigl((p\mli q)\adc (p\mli r)\bigr)\mli\bigr(p\mli (q\mlc r)\bigr)$. 
Indeed, this formula is instable, so it could only be derived by Rule {\bf (b)}. The premise of this rule should be 
either $(p\mli q)\mli\bigr(p\mli (q\mlc r)\bigr)$ or $(p\mli r)\mli\bigr(p\mli (q\mlc r)\bigr)$. In either case we deal with  a
formula that can be derived neither by Rule {\bf (a)} (because it is instable) nor by Rule {\bf (b)} (because it does not contain $\adc,\add$).
\end{example}

\begin{exercise}\label{ex2}
With $\mbox{\em Logic}\vdash F$ (resp. $\mbox{\em Logic}\not\vdash F$) here and later meaning ``$F$ is provable (resp. not provable) in {\em Logic}", show that:\vspace{3pt}

$\begin{array}{l} 
\mbox{$\propel\vdash \bigl((p\adc q)\mlc(p\adc q)\bigr)\mli (p\adc q)$};\\
\mbox{$\propel\not\vdash (p\adc q)\mli \bigl((p\adc q)\mlc(p\adc q)\bigr)$}.
\end{array}$
\end{exercise}

As we probably just had a chance to notice, if $F$ is an elementary formula, then the only way to prove $F$ in $\propel$ is to derive it by Rule {\bf (a)} from the empty set of premises. In particular, this rule will be applicable when $F$ is stable, which for an elementary $F$ means nothing but that $F$ is a classical tautology. And vice versa: every classically valid formula is an elementary formula derivable in $\propel$ by Rule {\bf (a)} from the empty set of premises. Thus we have:

\begin{proposition}\label{clas}
The $\adc,\add$-free fragment of $\propel$ is exactly classical propositional logic.
\end{proposition}

This is what we should have expected for, as noted earlier, 
when restricted to elementary problems --- and $\adc,\add$-free formulas are exactly the ones that represent such
 problems --- 
the meanings of $\gneg,\mlc,\mld,\mli$ are exactly classical. Here comes the soundness/completeness result: 

\begin{theorem} \label{main}
{\em (Japaridze \cite{cl1})} \ $\propel\vdash F$ iff $F$ is valid $($any $\propel$-formula $F$$)$.
\end{theorem}
 
Since the atoms of $\propel$ represent predicates rather than ICPs in general, $\propel$ only describes 
the valid principles of elementary ICPs. 
This limitation of expressive power is overcome in the extension of $\propel$ called $\propell$. The language of the latter augments the language of the former in that, along with the old atoms of $\propel$ which we now call {\bf elementary atoms}, it has an additional sort of (nonlogical) atoms called {\bf general atoms}. We continue using the lowercase letters 
$p,q,r,\ldots$  as metavariables for elementary atoms, and will be using the uppercase $P,Q,R,\ldots$ as metavariables for general atoms. We refer to formulas of this language as {\bf $\propell$-formulas}. An {\bf interpretation} now becomes a function that sends each nonlogical elementary atom (as before) to an elementary ICP and each general atom to any, not-necessarily-elementary, ICP. This mapping extends to all formulas in the same way as in the case of $\propel$. 
The concepts of validity, surface occurrence and positive/negative occurrence straightforwardly extend to 
this new language. The {\bf elementarization} of a $\propell$-formula $F$ means the result of replacing in $F$ every surface occurrence of the form $G\adc H$ by $\twg$, every surface occurrence of the form $G\add H$ by $\tlg$ {\em and}, in addition, replacing every positive surface occurrence of a general atom by $\tlg$ and every negative surface occurrence of a general atom by $\twg$. 

The rules of inference of $\propell$ are the two rules of $\propel$ --- that are now applied to any $\propell$-formulas rather than (just) $\propel$-formulas --- plus the following additional rule:

\begin{description}
\item[Rule (c):] $F'\mapsto F$, where $F'$ is the result of replacing in $F$ two --- one positive and one negative ---
surface occurrences of some general atom by a nonlogical elementary atom that does not occur in $F$.
\end{description}

\begin{example}\label{jan11} The following is a $\propell$-proof of $P\mlc P\mli P$:\vspace{3pt}

$\begin{array}{ll}
1.\ p\mlc P\mli p & \mbox{(from $\{\}$ by Rule {\bf (a)})}\\
2.\ P\mlc P\mli P & \mbox{(from 1 by Rule {\bf (c)})}
\end{array}$\vspace{3pt}

\noindent On the other hand, $\propell$ does not prove $P\mli P\mlc P$ (while, of course, it proves $p\mli p\mlc p$). Indeed, this formula is instable and does not contain $\adc$ or $\add$, so it cannot be derived by Rules {\bf (a)} or {\bf (b)}. If it is derived by Rule {\bf (c)}, the premise should be $p\mli P\mlc p$ or 
$p\mli p\mlc P$ for some elementary atom $p$. In either case we deal with an instable formula that contains no choice operators and only has one occurrence of a general atom, so that it cannot be derived by any of the three rules of $\propell$.
\end{example}

\begin{exercise} Verify that:\label{jan31}

1. $\propell\vdash P\mld\gneg P$

2. $\propell\not\vdash P\add\gneg P$

3. $\propell\vdash P\mli P\adc P$

4. $\propell\vdash (P\mlc Q)\mld (R\mlc S)\mli (P\mld R)\mlc(Q\mld S)$ \ (Blass's \cite{Bla92} principle)

5. $\propell\vdash p\mlc(p\mli Q)\mlc(p\mli R)\mli Q\mlc R$

6. $\propell\not\vdash P\mlc(P\mli Q)\mlc(P\mli R)\mli Q\mlc R$

7. $\propell\vdash P\adc(Q\mld R)\mli (P\adc Q)\mld (P\adc R)$

8. $\propell\not\vdash(P\adc Q)\mld (P\adc R)\mli P\adc(Q\mld R)$

9. $\propell\vdash(p\adc Q)\mld (p\adc R)\mli p\adc(Q\mld R)$
\end{exercise}

\begin{theorem}\label{thcl2} 
{\em (Japaridze \cite{cl2})} \ $\propell\vdash F$ iff $F$ is valid $($any $\propell$-formula $F$$)$.
\end{theorem}

Both $\propel$ and $\propell$ are obviously decidable, with a brute force decision algorithm running in polynomial space. 
Whether there are more efficient algorithms is unknown. 

A next step in exploring propositional computability logic would be augmenting the language of $\propel$ or $\propell$ with recurrence operators. At present the author sees the decidability 
of the set of valid formulas in the $\st,\pst$-augmented language of $\propel$, but has nothing yet to say about the $\st,\pst$-augmented $\propell$. 

\section{The first-order logic of computability}\label{sull}

$\propell$ seamlessly extends to the first-order logic $\predell$ with four quantifiers: $\cla,\cle,\ada,\ade$.
The set of {\em variables} of the language of $\predell$ is the same as the one that we fixed in Section 3. 
{\em Constants} $1,2,3,\ldots$ are also allowed in the language, and {\em terms} have the same meaning as before. 
The language has two --- {\bf elementary} and {\bf general} --- sorts of {\bf ICP letters},
where each such letter comes with a fixed integer $n\geq 0$ called its {\bf arity}. We assume that, for each $n$, there are infinitely many $n$-ary ICP letters of either (elementary and general) sort. 
Each {\bf atom} looks like $L(t_1,\ldots,t_n)$, where $L$ is an $n$-ary ICP letter and the $t_i$ are any terms. 
The terms ``elementary", ``general", ``$n$-ary"  extend from ICP letters to atoms in the obvious way. 
If $L$ is a $0$-ary ICP letter, then
the (only) corresponding atom we can write as $L$ rather than $L()$. 
$\twg$ and $\tlg$, as before, are two special ($0$-ary) elementary atoms called {\bf logical}. 

Formulas of this language, referred to as {\bf $\predell$-formulas}, are built from atoms using $\gneg$, $\mlc$, $\mld$, $\mli$, $\adc$, $\add$ in the same way as $\propel$- or $\propell$-formulas; in addition, we have the following formation rule:  If $F$ is a formula and $x$ is a variable, then $\cla x(F)$, $\cle x(F)$, $\ada x(F)$ and $\ade x(F)$ are formulas.  
   
An {\bf interpretation} for the language of $\predell$ is a function that sends each $n$-ary  
general (resp. elementary nonlogical) letter $L$ to 
an ICP (resp. elementary ICP) $L^*(x_1,\ldots,x_n)$, where the $x_i$ are pairwise distinct variables; in this case we say 
that $ ^*$ {\bf interprets} $L$ as $L^*(x_1,\ldots,x_n)$. 
Note that, just as in the propositional case, we do not insist that interpretations respect the arity of ICP letters.
Specifically, 
we do not require that the above $L^*(x_1,\ldots,x_n)$ depend on only (or all) the variables $x_1,\ldots,x_n$. Some caution is however necessary to avoid unpleasant collisions of variables, and also to guarantee that 
$\cla x$ and $\cle x$ are only applied to games for which they are defined, i.e. games that are unistructural in $x$. For this reason, we restrict interpretations to ``admissible" ones. For a $\predell$-formula $F$ and interpretation $ ^*$ we say that $ ^*$ is {\bf $F$-admissible} iff, for every $n$-ary ICP letter $L$ occurring in $F$,  where $ ^*$ interprets $L$ as $L^*(x_1,\ldots,x_n)$, the following two conditions are satisfied:
\begin{description}
\item[(i)] $L^*(x_1,\ldots,x_n)$ does not depend on any variables that are not among $x_1,\ldots,x_n$ but occur in $F$;
\item[(ii)] Suppose, for some terms $t_1,\ldots,t_n$ and some $i$ with $1\leq i\leq n$, $F$ has a subformula 
$\cla t_iG$ or $\cle t_iG$, where $G$ contains an occurrence of $L(t_1,\ldots,t_n)$ that is not in the scope (within $G$) 
of $\ada t_i$ or $\ade t_i$. Then $L^*(x_1,\ldots,x_n)$ is unistructural in $x_i$.
\end{description}
 The concept of admissible interpretation extends to any  set $S$ of $\predell$-formulas by stipulating that an interpretation $ ^*$ is  $S$-admissible iff it is $F$-admissible for every $F\in S$. Notice that 
condition (ii) is automatically satisfied for elementary ICP letters, because an elementary problem (i.e. $L^*(x_1,\ldots,x_n)$) is always unistructural. 
In most typical cases we will be interested in interpretations that interpret every $n$-ary ICP letter $L$ as a unistructural game $L^*(x_1,\ldots,x_n)$ that does not depend on any variables other than $x_1,\ldots,x_n$, so that both conditions (i) and (ii) will be automatically satisfied. With this remark in mind and in order to relax terminology, henceforth we will usually omit ``$F$-admissible" and simply say ``interpretation"; every time 
an expression $F^*$ is used in a context, it should be understood that the range of $^*$ is restricted to $F$-admissible interpretations.  

Every  interpretation $^*$ extends from ICP letters to formulas (for which $^*$ is admissible) in the obvious way: where $L$
 is an $n$-ary ICP letter interpreted as $L^*(x_1,\ldots,x_n)$ and $t_1,\ldots,t_n$ are any terms,
$\bigl(L(t_1,\ldots,t_n)\bigr)^*=L^*(t_1,\ldots,t_n)$; \ $\twg^*=\twg$, \ $(\gneg G)^*=\gneg (G^*)$; \ $(G\adc H)^*=(G^*)\adc (H^*)$; \ $(\cla xG)^*=\cla x(G^*)$, etc. When $F^*=A$, we say that $ ^*$ {\bf interprets} $F$ as $A$.
We say that a $\predell$-formula $F$ is {\bf valid} iff, for every ($F$-admissible) interpretation $^*$, the ICP $F^*$ is computable. 

The terms ``negative occurrence" and ``positive occurrence" have the same meaning as in the previous section. A {\bf surface occurrence} of a subexpression in a $\predell$-formula is an occurrence that is not in the scope of choice operators $\adc,\add,\ada,\ade$. When a $\predell$-formula contains neither choice operators nor general atoms, 
 it is said to be {\bf elementary}. The {\bf elementarization} of a $\predell$-formula $F$ is the result of replacing in $F$ 
every surface occurrence of the form $G\adc H$ or $\ada xG$ by $\twg$, every surface occurrence of the form $G\add H$ 
or $\ade  xG$ by $\tlg$, every positive surface occurrence of a general atom by $\tlg$ and every negative surface occurrence 
of a general atom by $\twg$. A $\predell$-formula is {\bf stable} iff its elementarization is a valid formula of classical first-order logic. The definition of a {\em free occurrence} of a variable $x$ in a formula is standard, meaning that 
the occurrence is not in the scope of $\cla x,\cle x,\ada x$ or $\ade x$. We will be using the expression $F(x/t)$ to denote 
the result of replacing all free occurrences of variable $x$ by term $t$ in $\predell$-formula $F$.
  
The rules of inference of $\predell$ are obtained from those of $\propell$ by replacing them by their ``first-order versions", with Rule {\bf (b)} splitting into two rules {\bf (B1)} and {\bf (B2)}, as follows:\vspace{-3pt}

\begin{description}
\item[Rule (A):]  $\vec{H}\mapsto F$, where $F$ is stable and $\vec{H}$ is 
a set of $\predell$-formulas satisfying the following conditions:\vspace{-2pt}
\begin{description}
\item[(i)]  Whenever $F$ has a positive (resp. negative)  surface occurrence of a subformula $G_1\adc G_2$ (resp. $G_1\add G_2$), for each 
$i\in\{1,2\}$, $\vec{H}$ contains the result of replacing that occurrence in $F$ by $G_i$;\vspace{-2pt}
\item[(ii)] Whenever $F$ has a positive (resp. negative) surface occurrence of a subformula $\ada x G$ (resp. $\ade xG$), $\vec{H}$ contains the result of replacing that occurrence in $F$ by $G(x/y)$ for some variable $y$ not occurring in $F$.\vspace{-2pt}
\end{description}
\item[Rule (B1):]  $F'\mapsto F$, where $F'$ is the result of replacing in $F$ a negative (resp. positive) surface occurrence of a subformula $G_1\adc G_2$ (resp. $G_1\add G_2$) by $G_i$ for some $i\in\{1,2\}$.\vspace{-2pt}
\item[Rule (B2):]  $F'\mapsto F$, where $F'$ is the result of replacing in $F$ a negative (resp. positive) surface occurrence of a subformula $\ada xG$ (resp. $\ade xG$) by $G(x/t)$ for some term $t$ such that (if $t$ is a variable) 
neither the above occurrence of $\ada xG$ (resp. $\ade xG$) within $F$ nor any of the free occurrences of $x$ within
$G$ are in the scope of $\cla t,\cle t,\ada t$ or $\ade t$.  
\item[Rule (C):] \  $F'\mapsto F$, where $F'$ is the result of replacing in $F$ two --- one positive and one negative ---
surface occurrences of some $n$-ary general ICP letter by an $n$-ary nonlogical elementary ICP letter that does not occur in $F$.
\end{description}

In what follows, the lowercase $p$ stands for a 1-ary (and hence nonlogical) elementary ICP letter, and the uppercase 
$P,Q$ for 1-ary general ICP letters.  
\begin{example} The following is a $\predell$-proof of $\ada x\ade y \bigl(P(x)\mli P(y)\bigr)$:\vspace{3pt}

$\begin{array}{ll}
1.\ \ p(z)\mli p(z) & \mbox{(from $\{\}$ by Rule {\bf (A)})}\\
2.\ \ P(z)\mli P(z) & \mbox{(from 1 by Rule {\bf (C)})}\\
3.\ \ \ade y\bigl(P(z)\mli P(y)\bigr) & \mbox{(from 2 by Rule {\bf (B2)})}\\
4.\ \ \ada x\ade y\bigl(P(x)\mli P(y)\bigr) & \mbox{(from \{3\} by Rule {\bf (A)})}
\end{array}$\vspace{3pt}

On the other hand, a little analysis can convince us that $\predell\not\vdash \ade y\ada x\bigl(P(x)\mli P(y)\bigr)$, even though the ``blind version" $\cle y\cla x\bigl(P(x)\mli P(y)\bigr)$ of this formula is derivable as follows:\vspace{3pt}

$\begin{array}{ll}
1.\ \ \cle y\cla x \bigl(p(x)\mli p(y)\bigr) & \mbox{(from $\{\}$ by Rule {\bf (A)})}\\
2.\ \ \cle y\cla x \bigl(P(x)\mli P(y)\bigr) & \mbox{(from 1 by Rule {\bf (C)})}
\end{array}$

\end{example}

\begin{exercise}\label{Jan20} Verify that:\vspace{3pt}

$\begin{array}{l}
1.\ \ \predell\vdash \cla xP(x)\mli \ada xP(x)\\
2.\ \ \predell\not\vdash \ada xP(x)\mli \cla xP(x)\\
3.\ \ \predell\vdash \ada x\bigl(\bigl(P(x)\mlc\ada xQ(x)\bigr)\adc\bigl(\ada xP(x)\mlc Q(x)\bigr)\bigr)\mli \ada xP(x)\mlc \ada xQ(x) 
\end{array}$
\end{exercise}

A little excursus for the logicians. It was noted in Section 1 that the logical behavior of our parallel and choice 
operators is similar to yet by no means the same as that of the ``corresponding" multiplicative and additive operators of 
linear logic (LL). Now we can be more specific. CL and LL agree on many simple and demonstrative formulas such as $P\mli 
P\mlc P$ and $P\add\gneg P$ that both logics reject (Example \ref{jan11}, Exercise \ref{jan31}), or $P\mld\gneg P$ and $P\mli P\adc P$ that both logics accept (Exercise \ref{jan31}). CL also agrees with the version of LL called affine logic (LL with the weakening rule) on $P\mlc P\mli P$ that both logics accept. 
On the other hand, the somewhat longer formulas of Exercises \ref{Jan20}(3) and \ref{jan31}(4) are valid in our sense yet underivable in linear (or affine) logic. 
Neither the similarities nor the discrepancies are a surprise. The philosophies of CL and LL overlap in their striving to develop a logic of resources. But the ways this 
philosophy is materialized are radically 
opposite. CL starts with a mathematically strict and intuitively convincing semantics, and only after that, as a natural second step, 
asks what the corresponding logic and its axiomatizations (syntax) are. 
LL, on the other hand, started directly from the second step. It was introduced syntactically rather than semantically, essentially by taking classical sequent calculus and throwing out the rules that seemed to be unacceptable from the intuitive, naive resource point of view, so that, in the  
absence of a clear concept of truth or validity, the question about whether the resulting system was sound/complete could not even be meaningfully asked. In this process of syntactically rewriting classical logic some innocent, deeply hidden principles could have easily gotten victimized. Apparently the above-mentioned formulas separating CL from LL should be considered examples of such ``victims". 
Of course, many attempts have been made to retroactively find a missing semantical justification for LL.
Technically it is always possible to come up with some sort of a formal semantics that matches a given target syntactic construction, but the whole question is how natural and meaningful such a semantics is in its own right, and how adequately it captures the logic's underlying philosophy and ambitions. Unless, by good luck, the target 
system really {\em is} ``the right logic", the chances of a decisive success when following the odd scheme `from syntax to semantics' can be rather slim. The natural scheme is `from semantics to syntax'. It matches the way classical logic evolved and climaxed in G\"{o}del's completeness theorem. And this is exactly the scheme that CL, too, follows.      

Taking into account that classical validity and hence stability is recursively enumerable, obviously (the set of theorems of) {\em $\predell$ is recursively enumerable}. 
 \cite{cl5} also proves that  

\begin{theorem} The $\cla,\cle$-free fragment of $\predell$ is decidable.
\end{theorem}

This is a nice and perhaps not very obvious/expected fact, taking into account that the above fragment of $\predell$  is still a first order logic as it contains the quantifiers $\ada$,$\ade$. This fragment is also natural as it gets rid of the only operators of the language that produce games with imperfect information. 

Next, based on the straightforward 
observation that elementary formulas are derivable in  $\predell$ (in particular, from $\{\}$ by Rule {\bf (A)}) 
exactly when they are classically valid, we have: 

\begin{proposition} $\predell$ is a conservative extension of classical first-order logic: an elementary  $\predell$-formula is classically valid if and only if it is provable in $\predell$. 
\end{proposition}

The following theorem is the strongest result on CL known so far, with the aggregate size of its formal proof 
exceeding 70 pages:
 

\begin{theorem}\label{main5} 
{\em (Japaridze \cite{cl5})} \ $\predell\vdash F$ iff $F$ is valid $($any $\predell$-formula $F$$)$.
Furthermore:

{\bf Uniform-Constructive Soundness:} There is an effective  procedure that takes a $\predell$-proof of an arbitrary formula $F$ and 
constructs an HPM $\cal M$ such that,  for every \mbox{interpretation $^*$,} \ $\cal M$ solves $F^*$.

{\bf Strong Completeness:} If $\predell\not\vdash F$, then $F^*$ is not computable  for some \mbox{interpretation $^*$} that interprets \ each elementary atom as a finitary predicate \ and \ each general atom as  a \mbox{$\adc,\add$-combination} of 
finitary predicates.  
\end{theorem}

Here ``{\bf finitary predicate}" (or {\bf finitary game} in general) is a predicate (game) $A$ for which there is some finite set $X$ of variables such that, for any two valuations $e_1$ and $e_2$ that agree on all variables from $X$, we have $e_1[A]=e_2[A]$.
That is, only the values of those finitely many variables are relevant. A non-finitary game generally depends on infinitely many variables, and appealing to this sort of games in a completeness proof could seriously weaken such a result: the reason for incomputability of a non-finitary game could be just the fact that the machine can never finish reading all the relevant information from its valuation tape. Fortunately, in view 
of the Strong Completeness clause, it turns out that the question whether non-finitary ICPs are allowed or not has no effect on the soundness and completeness of $\predell$; moreover, ICPs can be further restricted to the sort of 
games as simple as $\adc,\add$-combinations of finitary predicates. Similarly, the Uniform-Constructive Soundness clause dramatically strengthens the soundness result for $\predell$ and, as this will be discussed in Section 8, opens application areas far beyond the pure theory of computing. Of course, both uniform-constructive soundness and strong completeness (automatically) hold for $\propel$ and $\propell$ as well, but the author has chosen to disclose this good news only in the
present section. 

Theorem \ref{main5}, even though by an order of magnitude more informative than G\"{o}del's completeness theorem for classical logic 
which it implies as a special case, 
is probably only a beginning of progress on the way of in-depth study of computability logic. Seeing what happens if we add parallel quantifiers and/or the recurrence group of operators to the language of $\predell$, or exploring some other --- limited --- $\mla,\st,\pst$-containing fragments of CL,
remains a challenging but worthy task to pursue. Among the interesting fragments of CL is the one that only has general atoms and the operators $\adc,\add,\ada,\ade,\intimpl$, where $A\intimpl B$ is defined as $(\st A)\mli B$. It was conjectured in \cite{Jap03} that the valid formulas of this language are exactly those provable in Heyting's intuitionistic calculus, with the above operators understood as the intuitionistic conjunction, disjunction, universal quantifier, existential quantifier and implication, respectively. The soundness part of this conjecture was successfully verified 
later in \cite{Japint}. A verification of the remaining completeness part of the conjencture could signify a convincing ``proof" of Kolmogorov's (1932) well-known but so 
far rather abstract thesis according to which intuitionistic logic is a logic of problems. 


\section{Applied systems based on CL}\label{applc}

The original motivation underlying CL, presented in Section 1, was computability-theoretic: the approach provides a systematic  
answer to the question `what can be computed?', which is a fundamental question of computer science. Yet, a  look at the Uniform-Constuctive Soundness clause of Theorem \ref{main5} 
reveals that the CL paradigm is not only about {\em what} can be computed. It is equally about 
{\em how} problems can be computed/solved, suggesting that CL should have substantial utility, with its application areas   
not limited to theory of computing. In the present section we will briefly examine why 
and how CL is of interest in some other fields of study, such as knowledgebase systems, planning systems or constructive applied theories.

The reason for the failure of $p\add\neg p$ as a computability-theoretic principle is  
that the problem represented by this formula may have no effective solution --- that is, the predicate $p^*$ may be undecidable. The reason why this principle fails in the context of knowledgebase systems, however, is much simpler. A knowledgebase system may fail to solve the problem
$\mbox{\em Female}(\mbox{\em Dana})  \add \gneg\mbox{\em Female}(\mbox{\em Dana})$ not because the latter has no effective solution (of course it has one), but because the system simply lacks sufficient knowledge to determine Dana's gender. On the other hand, any system would be able to ``solve" the problem
 $\mbox{\em Female}(\mbox{\em Dana})  \mld   \gneg\mbox{\em Female}(\mbox{\em Dana})$
as this is an automatically won elementary game so that there is nothing to solve at all. 
Similarly, while $\cla y\cle x\mbox{\em Father}(x,y)$ is an automatically solved elementary problem expressing the almost tautological knowledge that every person has a father, ability to solve the problem  
$\ada y\ade x\mbox{\em Father}(x,y)$ implies the nontrivial knowledge of everyone's actual father. 
Obviously the knowledge expressed by $A\add B$ or $\ade xA(x)$ is generally stronger than the knowledge expressed by $A\mld B$ or $\cle xA(x)$, yet the language of classical logic fails to capture this difference --- the difference whose relevance hardly requires any explanation. The traditional approaches to knowledgebase systems 
(\cite{Kon89,Moo85} etc.) 
try to mend this gap by augmenting the language of classical logic with special epistemic constructs, such as the 
 modal ``know that" operator $\Box$, after which probably $\Box A
 \mld \Box B$ would be suggested as a translation for $A \add B$  and $\cla y\cle x \Box A(x,y)$ for $\ada y\ade xA(x,y)$. Leaving it for the philosophers to argue whether,
say, $\cla y\cle x \Box A(x,y)$ really expresses the constructive meaning 
of $\ada y\ade xA(x,y)$, and forgetting that epistemic constructs typically yield unnecessary and very unpleasant  complications such as messiness and non-semidecidability of the resulting logics, some of the major issues still do not seem to be taken care of. Most of the actual knowledgebase and information systems are interactive, and what we really need is a logic of {\em interaction} rather than just a logic 
of knowledge. Furthermore, a knowledgebase logic needs to be {\em resource-conscious}. The informational resource
expressed by $\ada x(\mbox{\em Female}(x)  \add   \gneg\mbox{\em Female}(x))$ is not as strong as
the one expressed by  $\ada x(\mbox{\em Female}(x)  \add   \gneg\mbox{\em Female}(x))\mlc\ada x(\mbox{\em Female}(x)  \add   \gneg\mbox{\em Female}(x))$: the former implies the resource provider's commitment to tell only one (even though an arbitrary one) person's gender, while the latter is about telling any two people's genders.\footnote{A reader having difficulty in understanding why this difference is relevant, may try to replace {\em Female}$(x)$ with {\em Acid}$(x)$, and then think of a (single) 
piece of litmus paper.} Neither classical logic nor its standard epistemic extensions have the ability to account for such differences. CL promises to be adequate. 
It {\em is} a logic of interaction, it {\em is} resource-conscious, and it {\em does} 
capture the relevant differences between truth and actual ability to find/compute/know truth.  
 
When CL is used as a logic of knowledgebases, its formulas represent interactive queries. A formula whose main operator is $\add$ or $\ade$ can be understood as a question asked by the user, and a formula whose main operator is $\adc$ or $\ada$  a question asked by the system.
Consider the problem $\ada x \ade y \mbox{\em Has}(x, y)$, where $\mbox{\em Has}(x,y)$ means ``patient $x$ has disease  $y$" (with {\em Healthy} counting as one of the possible ``diseases"). This formula is the following question asked by the system: ``Who do you want me to diagnose?"  The user's response can be ``Dana". This move brings the game down to 
$\ade y \mbox{\em Has}(Dana, y)$. This is now a question asked by the user: ``What does Dana have?". The system's response can be ``flu", taking us to the terminal position \(\mbox{\em Has}(Dana,\- Flu).\) The system has been successful iff Dana really has flu. 

Successfully solving the above problem $\ada x \ade y \mbox{\em Has}(x, y)$ requires having all relevant medical information for each possible patient, which in a real diagnostic system would hardly be the case. Most likely, such a system, after
receiving a request to diagnose $x$, would make counterqueries regarding $x$'s symptoms, blood pressure, test results, age, gender, etc., so that the query that the system will be solving would have a higher degree of interactivity than the two-step query $\ada x \ade y \mbox{\em Has}(x, y)$ does, with questions and counterquestions interspersed in some complex fashion. Here is when other computability-logic operations come into play. 
$\gneg$ turns queries into counterqueries; parallel operations generate combined queries, with $\mli$ acting as a query reduction operation; $\st,\pst$ allow repeated queries, etc. Here we are expanding our example. 
Let {\em Symptoms}$(x,s)$ mean ``patient $x$ has (set of) symptoms $s$", and $Positive(x,t)$ mean ``patient $x$ tests 
positive for test $t$". Imagine a diagnostic system that can diagnose any particular patient $x$, but needs some additional information. Specifically, it needs to know $x$'s symptoms; plus, the system may require to have $x$ taken a test $t$ that it selects dynamically in the course of a dialogue with the user depending on what responses it received. The interactive task/query that such a system is performing/solving can then be expressed by the formula\vspace{-3pt} 
\begin{equation}\label{diagn}\ada x\Bigl(\ade s\mbox{\em Symptoms}(x,s)\mlc \ada t\bigl(\mbox{\em Positive}(x,t)\add\gneg \mbox{\em Positive}(x,t)\bigr)\mli \ade y\mbox{\em Has}(x,y)\Bigr).\vspace{-3pt}\end{equation}
A possible scenario of playing the above game is the following. At the beginning, the system waits until the user specifies a patient $x$ to be diagnosed. We can think of this stage as systems's requesting the user to select a particular (value of) $x$, remembering that the presence of $\ada x$ automatically implies such a request. 
After a patient $x$ --- say $x=X$ --- is selected,
the system requests to specify $X$'s symptoms. Notice that our game rules 
make the system successful if the user fails to provide this information, i.e. specify a (the true) value for $s$ in 
$\ade sSymptoms(X,s)$. 
Once a response --- say, $s=S$ --- is received, the system selects a test $t=T$ and asks the user to perform it on $X$, i.e. to choose the true disjunct of $\mbox{\em Positive}(X,T)\add\gneg \mbox{\em Positive}(X,T)$. Finally, provided that the user gave correct answers to all counterqueries 
(and if not, the user has lost), the system
makes a diagnostic decision, i.e. specifies a value $Y$ for $y$ in   $\ade y\mbox{\em Has}(X,y)$ for which 
$\mbox{\em Has}(X,Y)$ is true. 

The presence of a single ``copy" of $\ada t\bigl(\mbox{\em Positive}(x,t)\add\gneg \mbox{\em Positive}(x,t)\bigr)$
in the antecedent of (\ref{diagn})  means that the system may request testing a given patient only once. If $n$ tests were potentially needed instead, 
this would be expressed by taking the $\mlc$-conjunction of $n$ identical conjuncts $\ada t\bigl(\mbox{\em Positive}(x,t)\add\gneg \mbox{\em Positive}(x,t)\bigr)$. 
And if the system potentially needed 
an unbounded number of tests, then we would write $\pst\ada t\bigl(\mbox{\em Positive}(x,t)\add\gneg \mbox{\em Positive}(x,t)\bigr)$, thus further weakening (\ref{diagn}): a system that performs this weakened task is not as good as the one performing (\ref{diagn}) as it requires stronger external (user-provided) informational resources. Replacing the main 
quantifier $\ada x$ by $\cla x$, on the other hand, would strengthen (\ref{diagn}), signifying the system's ability 
to diagnose a patent purely on the basis of his/her symptoms and test result without knowing who the patient really is. However, if in its diagnostic decisions the system uses some additional information on patients such 
their medical histories stored in its knowledgebase and hence needs to know the patient's identity,
$\ada x$ cannot be upgraded to $\cla x$. Replacing $\ada x $ by $\mla x$ would be a yet another way to strengthen (\ref{diagn}),
signifying the system's ability to diagnose 
all patients rather than any particular one; obviously effects of at least the same strength would be achieved by 
just prefixing (\ref{diagn}) with $\pst$ or $\st$. 

As we just mentioned system's {\bf knowledgebase}, let us make clear what it means. Formally, this is a finite 
$\mlc$-conjunction {\em KB}
of formulas, which can also be thought of as the (multi)set of its conjuncts. We call the elements of this set the {\bf internal informational resources} of the system. Intuitively, {\em KB} represents all of the 
nonlogical knowledge available to the system, so that (with a fixed built-in logic in mind) the strength of the former 
determines the query-solving power of the latter. Conceptually, however, we do not think of {\em KB} as a 
part of the system properly. The latter is just ``pure", logic-based problem-solving software of universal utility that 
initially comes to the user without any nonlogical knowledge whatsoever. Indeed, built-in nonlogical knowledge 
would make it no longer universally applicable: Dana can be a female in the world of one potential user while a male in the 
world of another user, and $\forall x\forall y(x\times y=y\times x)$ can be false to a user who understands $\times$ as Cartesian rather than number-theoretic product. It is the user who selects and maintains {\em KB} for the system, putting into it all informational resources that (s)he believes are relevant, correct and maintainable. Think of the formalism of CL as a highly declarative programming language, and the process of creating {\em KB} as programming in it.

The knowledgebase {\em KB} of the system may include atomic elementary formulas expressing factual knowledge, such as $\mbox{\em Female}(\mbox{\em Dana})$, or non-atomic elementary formulas expressing general knowledge, such as $\cla x
\bigl(\cle y\mbox{\em Father}(x,y)\mli\mbox{\em Male}(x)\bigr)$ or $\cla x\cla 
y\bigl(x\times(y+1)=(x\times y)+x\bigr)$; it can also include nonelementary formulas such as 
$\st\ada x\bigl({\em Female}(x)\add\mbox{\em
 Male}(x)\bigr)$, expressing potential knowledge of everyone's gender, or $\st\ada x\ade y(x^2=y)$, 
expressing ability to repeatedly compute the square function, or something more complex and more interactive such as formula (\ref{diagn}). With each resource $R\in${\em KB} is associated (if not physically, at least conceptually) its {\bf provider} --- an agent that solves the query $R$ for the system, i.e. plays the game $R$ against the system. 
Physically the provider could be a computer program allocated to the system, or a network server having the system 
as a client, or another knowledgebase system to which the system has querying access, or even human personnel servicing the system. E.g., the provider for $\st\ada x\ade y\mbox{\em Bloodpressure}(x,y)$ would probably be a team of nurses repeatedly performing the task of measuring the blood pressure of a patient specified by the system and reporting the outcome back to the system. Again, we do not think of providers as a part of the system itself. The latter only sees {\em what} resources are available to it, without knowing or caring about {\em how} the corresponding providers do their job; furthermore,
the system does not even care {\em whether} the providers really do their job right.   
The system's responsibility is only to correctly solve queries for the user {\em as long as} none of the providers 
fail to do their job.  Indeed, if the system misdiagnoses a patient because a nurse-provider gave it wrong information 
about that patient's blood pressure, the hospital (ultimate user) is unlikely to  fire the system and demand refund from its vendor; more likely, it would fire the nurse. 
Of course, when $R$ is elementary, the 
provider has nothing to do, and its successfully playing $R$ against the system simply means that $R$ is true. 
Note that in the picture that we have just presented, the system plays each game $R\in${\em KB} in the role of $\oo$, so that, from the system's perspective, the game that it plays against the provider of $R$ is $\gneg R$ rather than $R$.

The most typical internal informational resources, such as factual knowledge or queries solved by computer programs, can be reused an arbitrary number of times and with unlimited branching capabilities, i.e. in the strong sense captured by the operator $\st$, and thus they would be prefixed with $\st$ as we did with $\ada x\bigl({\em Female}(x)\add\mbox{\em
 Male}(x)\bigr)$ and $\ada x\ade y(x^2=y)$. There was 
no point in $\st$-prefixing $\mbox{\em Female}(\mbox{\em Dana})$, $\cla x
\bigl(\cle y\mbox{\em Father}(x,y)\mli\mbox{\em Male}(x)\bigr)$  or $\cla x\cla y\bigl(x\times(y+1)=(x\times 
y)+x\bigr)$ because every elementary game $A$ is equivalent to $\st A$ and hence remains ``recyclable" even without recurrence operators. As this was noted in Section 1, there is no difference between $\st$ and $\pst$ as long as ``simple" 
resources such as $\ada x\ade y(x^2=y)$ are concerned. However, 
in some cases --- say, when a resource with a high degree of interactivity is supported by an unlimited number of independent providers each of which however allows to run only one single ``session"  --- the weaker operator $\pst$ will have to be used instead of $\st$. 
Yet, some of the internal informational resources could be essentially non-reusable. A provider possessing 
a single item of disposable pregnancy test device would apparently be able to support the resource  $\ada x(\mbox{\em Pregnant}(x)\add\gneg \mbox{\em Pregnant}(x)\bigr)$ but not  $\st \ada x(\mbox{\em Pregnant}(x)\add\gneg \mbox{\em Pregnant}(x)\bigr)$
and not even $\ada x(\mbox{\em Pregnant}(x)\add\gneg \mbox{\em Pregnant}(x)\bigr)\mlc \ada x(\mbox{\em Pregnant}(x)\add\gneg \mbox{\em Pregnant}(x)\bigr)$. Most users, however, would try to refrain from including this sort of a resource into 
{\em KB} but rather make it a part (antecedent) of possible queries. Indeed, knowledgebases with non-recyclable resources would tend to weaken from query to query and require more careful maintainance/updates. The appeal 
of a knowledgebase entirely consisting of $\st$,$\pst$-resources is its absolute persistence.
Whether recyclable or not, all of the resources of {\em KB} can be used independently and in parallel. This is exactly what allows us to identify {\em KB} with the $\mlc$-conjunction of its elements. 

Assume {\em KB}\hspace{2pt}$=R_1\mlc\ldots\mlc R_n$, and let us now try to visualize the picture of the system solving a query $F$ for the user. The designer would probably select an interface where the user only sees the moves made by the system in 
$F$, and hence gets the illusion that the system is just playing $F$. But in fact the game that the system is really playing 
is {\em KB}\hspace{2pt}$\mli F$, i.e. $\gneg R_1\mld\ldots\mld\gneg R_n\mld F$. Indeed, the system is not only interacting with the user 
in $F$, but --- in parallel --- also with its providers against whom, as we already know, it plays $\gneg R_1,\ldots,\gneg R_n$. As long as those providers do not fail to do their job, the system loses each of  
the games $\gneg R_1,\ldots,\gneg R_n$. Then our semantics for $\mld$ implies that the system wins its play over  
the ``big game" $\gneg R_1\mld\ldots\mld\gneg R_n\mld F$ if and only if it wins it in the $F$ component, i.e. 
successfully solves the query $F$. 

Thus, the system's ability to solve a query $F$ reduces to its ability to generate a solution to {\em KB}\hspace{2pt}$\mli F$, i.e. a reduction of $F$ to {\em KB}. What would give the system such an ability is built-in knowledge of CL --- in particular, a {\bf uniform-constructively sound axiomatization}
of it, by which we mean a deductive system $S$ (with effective proofs of its theorems) that satisfies the Uniform-Constructive Soundness clause of Theorem \ref{main5} with ``$S$" in the role of $\predell$. According to the uniform-constructive soundness property, it would be sufficient for the system to find a proof of {\em KB}$\mli F$, which would allow it to (effectively) construct an HPM $\cal M$ and then run it on {\em KB}$\mli F$ with guaranteed success. 
  
Notice that it is uniform-constructive soundness rather than simple soundness of the 
the built-in (axiomatization of the) logic that allows the knowledgebase system to function. Simple soundness just means that every provable formula is valid. This is not sufficient for two reasons. One reason is that validity of a formula $E$ only implies that, for every interpretation $^*$, a solution to the problem $E^*$ exists. It may be the case, however, that 
different interpretations require different solutions, so that choosing the right solution requires knowledge of the 
actual interpretation, i.e. the {\em meaning}, of the atoms of $E$. Our assumption is that the system has no 
nonlogical knowledge which, in more precise terms, means nothing but   that it has no knowledge of the interpretation $^*$. Thus, a solution that the system 
generates for {\em KB}\hspace{1pt}$^*\mli F^*$ should be successful for any possible interpretation $^*$. We call such an interpretation-independent  solution --- 
an HPM $\cal M$ that wins $E^*$ for every interpretation $^*$ --- a {\bf uniform solution} to $E$, and correspondingly call a formula {\bf uniformly valid} iff it has a uniform solution. The Uniform-Constructive Soundness clause asserts that every provable formula is not only valid, but also uniformly valid. Going back to the example with which this section started, the reason why 
$p\add\gneg p$ fails in the context of computability theory is that it is not valid, while the reason for the failure 
of this principle in the context of knowledgebase systems is that it is not uniformly valid: its solution, even if it existed for each interpretation $^*$, generally would depend on whether $p^*$ is true or false, and the system would  be unable to 
figure out the truth status of $p^*$ unless this information was explicitly or implicitly contained in {\em KB}. Thus, for knowledgebase systems the primary semantical concept of interest is uniform validity rather than validity.     
But does having two different concepts of validity mean that we will have to deal with two different logics? Not really. According to Conjecture 26.2 of \cite{Jap03}, {\em a formula of the language of CL is valid if and only if it is uniformly valid}. Our Theorem \ref{main5} with its Uniform-Constructive  Soundness clause signifies a successful verification of this conjecture for $\predell$-formulas: such a formula is valid iff it is uniformly valid iff it is provable in $\predell$. 
There are good reasons to expect that this nice extensional equivalence between validity and uniform validity continues
to hold for all reasonable extensions of the language of $\predell$ and, in particular, its extension with 
$\st,\cost,\pst,\pcost,\mla,\mle$.   

The other reason why simple soundness of the built-in logic would not be sufficient for a knowledgebase system to 
function --- even if every provable formula was known to be uniformly valid --- is the following. With simple soundness, after finding a proof of $E$, even though the system would know that a solution to $E^*$ exists, it might have no way to actually find such a solution. On the other hand, uniform-constructive soundness guarantees that a (uniform) solution to every provable formula not only exists, but can be effectively extracted from a proof.

As for completeness of the built-in logic  --- unlike uniform-constructive soundness --- it is a desirable but not necessary condition. So far a complete axiomatization has been 
found only for the fragment of CL limited to the language of $\predell$. We hope that the future will bring completeness results for more expressive fragments as well. But even if not, we can still certainly 
succeed in finding ever stronger axiomatizations that are uniform-constructively sound even if not necessarily complete. 
Extending $\predell$ with some straightforward rules such as the ones that allow to replace $\st F$ by $F\mlc \st F$ and $\pst F$ by $F\mlc\pst F$, the rules $F\mapsto \st F$, $F\mapsto \pst F$, etc. 
would already immensely strengthen the logic. It should also be remembered that, when it 
comes to practical applications in the proper sense, the logic that will be used is likely to be far from complete anyway. Say, the popular classical-logic-based systems and programming languages are incomplete, and the reason is not that 
a complete axiomatization for classical logic is not known, but rather the unfortunate fact of life that often 
efficiency only comes at the expense of completeness. 

But even $\predell$, as it is now, is already very powerful. Why don't we see a simple example to feel 
the flavor of it as a query-solving logic. Let {\em Acid}$(x)$ mean 
``solution $x$ contains acid", and {\em Red}$(x)$ mean ``litmus paper turns red in solution $x$". 
Assume that the knowledgebase {\em KB} of a $\predell$-based system contains $
\cla x\bigl(\mbox{\em Red}(x)\mli\mbox{\em Acid}(x)\bigr)$, $\cla x\bigl(\mbox{\em Acid}(x)\mli\mbox{\em Red}(x)\bigr)$ and $\ada x\bigl(\mbox{\em Red}(x)\add\gneg \mbox{\em Red}(x)\bigr)$, accounting for knowledge of the fact that a solution contains acid iff the litmus paper turns red in it, and for  
availability of a provider who possesses a piece of litmus paper that it can dip into any solution and report the paper's color to the system. 
Then the system 
can solve the acidity query $\ada x\bigr(\mbox{\em Acid}(x)\add\gneg\mbox{\em Acid}(x)\bigr)$. 
This follows from the fact --- left as an exercise for the reader to verify --- that $\predell\vdash \mbox{\em KB}\mli \ada x\bigr(\mbox{\em Acid}(x)\add\gneg\mbox{\em Acid}(x)\bigr)$.
 
An implicit assumption underlying our discussions so far was that an interpretation    
is fixed in a context and does change its values.  
Making just one more step and departing from this unchanging-interpretation assumption 
opens significantly wider application areas for CL, in particular, the more general area of planning and physical-informational (vs. just informational) resource management systems. We call such (CL-based) systems 
{\bf resourcebase systems}. In this new context, interpretations in the old, unchanging sense can be renamed into 
{\bf situations}, with the term ``interpretation" reserved for the more general concept of possibly dynamic mapping 
from atoms to ICPs --- mapping whose values may keep changing from situation to situation, 
with situations intuitively being nothing but ``snapshots" of interpretations. 
Dynamic interpretations are indeed the common case in real world. Perhaps Dana is not pregnant in a given situation, so that $(${\em Pregnant}$(${\em Dana}$))^*=\tlg$. But it may 
happen that  
the situation changes so that $^*$  reinterprets {\em Pregnant}$(${\em Dana}$)$ into $\pp$. Furthermore, probably 
Dana has full control over whether she gets pregnant or not. This means that she can successfully maintain 
the resource {\em Pregnant}$(${\em Dana}$)\adc\gneg${\em Pregnant}$(${\em Dana}$)$ which --- unlike {\em Pregnant}$(${\em Dana}$)\add\gneg${\em Pregnant}$(${\em Dana}$)$ --- generally no agent would be able to maintain if the situation 
was fixed and unmanageable.  Thus, in the context of resourcebase systems, successful game-playing no longer means just correctly answering 
questions. It may involve performing physical tasks, i.e. controlling/managing situations. Think of the 
task performed by a ballistic missile. With $t$ ranging over all reachable targets, this task can be expressed by $\ada t\mbox{\em Destroyed}(t)$. The user makes a move by specifying a target $t=T$. This amounts to commanding the 
missile to destroy $T$. Provided that the latter indeed successfully performs its task, the user's command will be satisfied: the situation, in which (the interpretation of) $\mbox{\em Destroyed}(T)$ was probably false, will change and 
 $\mbox{\em Destroyed}(T)$ become true. The same example demonstrates the necessity for a planning logic to be resource-conscious. With only one missile available as a resource, an agent would be able to destroy any one target but not two. 
This is accounted for by the fact that $\ada t\mbox{\em Destroyed}(t)\mli\mbox{\em Destroyed}(x)$ is valid 
while $\ada t\mbox{\em Destroyed}(t)\mli\mbox{\em Destroyed}(x)\mlc\mbox{\em Destroyed}(y)$ is not. 

The earlier-discussed CL-based knowledgebase systems solve problems in a uniform,  
interpretation-independent way. This means that whether the interpretation is unchanging or dynamic is technically 
irrelevant for them, so that exactly the same systems, without any modifications whatsoever, can be 
used for solving planning problems (instead of just solving queries) such as how to destroy target $T$ or how to make Dana
pregnant, with their knowledgebases ({\em KB}) --- renamed into {\bf resourcebases} ({\em RB}) --- now containing physical, situation-managing resources such as $\ada t\mbox{\em Destroyed}(t)$ along with old-fashioned informational resources.
See Section 26 of \cite{Jap03} for an illustrative example of a planning problem
solved with CL. CL and especially extensions of its present version with certain new game operators, such as sequential versions of conjunction/disjunction, quantifiers and recurrence operators,\footnote{Here is an informal outline of one of 
the --- perhaps what could be called {\em oblivious} --- versions of sequential operators. The {\bf sequential conjunction} $A\scc B$ is a game that starts and proceeds as a play of $A$; it will also end as an ordinary play of $A$ unless, at some point, $\oo$ makes a special ``switch" move; to this move --- it is OK if with a delay --- $\pp$ should respond  
with an ``acknowledgment" move (if such a response is never made, $\pp$ loses), 
after which $A$ is abandoned, and the play continues/restarts as a play of $B$ without the possibility to go back to $A$. The {\bf sequential universal quantification} $\sca xA(x)$ is then defined as 
$A(1)\scc A(2)\scc A(3)\scc\ldots$, and the {\bf sequential recurrence} $\ssti A$ as 
$A\scc A\scc A\scc\ldots$. 
In both cases $\oo$ is considered the loser if it makes a switch move infinitely many times. 
As this can be understood, the dual operators: {\bf sequential disjunction} $\scd$, {\bf sequential existential quantifier} $\sce$ and {\bf sequential corecurrence} $\scosti$ will be defined in a symmetric way with the roles of the two players interchanged.
 Note that, as a resource, $\ssti A$ is the weakest among 
$\ssti A,\psti A,\sti A$: just like $\psti A$ and $\sti A$, $\ssti A$ allows the user to restart $A$ an arbitrary number of times;
however, unlike the case with $\psti A$ and $\sti A$, only one session of $A$ can be run at a time, and restarting $A$ signifies giving up the previous run(s) of it.}  
might have good potential as a new logical paradigm for AI planning systems. 

The fact that CL is a conservative extension of classical logic also makes it a reasonable and appealing alternative to the latter in its most traditional and unchallenged application areas. In particular, it makes perfect sense to base applied theories --- such as, say, Peano arithmetic (axiomatic number theory) --- on CL instead of classical logic. Due to conservativity, no old informations 
would be lost or weakened this way. On the other hand, we would get by an order of magnitude more expressive, constructive and computationally meaningful theories than their classical-logic-based versions. Let us see a little more precisely 
what we mean by a CL-based applied theory. 
For simplicity, we restrict our considerations to the cases when the set {\em AX} of nonlogical {\bf axioms} of the
applied theory is finite. As we did with {\em KB}, we identify {\em AX} with the $\mlc$-conjunction of its elements. 
From (the problem represented by) {\em AX} --- or, equivalently, each conjunct of it --- we require to be computable in our sense, i.e. come with an HPM that solves it. So, notice, all of the axioms of the old, classical-logic-based version of the theory could be automatically included into the new set {\em AX} because they represent true and hence computable elementary problems. Many of those old axioms can be constructivized by, say, replacing blind or parallel operators with their choice equivalents. E.g., we would want to rewrite the axiom $\cla x\cle y(y=x+1)$ of arithmetic as the more informative $\ada x\ade y(y=x+1)$. And, of course, to the old axioms or their constructivized versions could be added some essentially new axioms expressing basic computability principles specific to (the particular interpretation underlying) the theory. Provability (theoremhood) of a formula $F$ in such a  
theory we understand as provability of the formula {\em AX}$\mli F$ in the underlying axiomatization of CL which, 
as in the case of knowledgebase systems, is assumed to be uniform-constructively sound. The rule of modus ponens 
has been shown in \cite{Jap03} (Proposition 21.3) to preserve computability in the following uniform-constructive sense:

\begin{theorem} There is an effective  function  \ $f$: \{HPMs\}$\times$\{HPMs\} $\rightarrow$ \{HPMs\} such that, for any HPMs $\cal M$,$\cal N$ and ICPs 
$A$,$B$, if $\cal M$ solves $A$ and $\cal N$ solves $A\mli B$, then $f({\cal M},{\cal N})$ solves $B$.  
\end{theorem}

\noindent This theorem, together with our assumptions that {\em AX} is computable and that the underlying logic is 
uniform-constructively sound, immediately implies that the problem represented by any theorem $F$ of the applied theory is computable and that, furthermore, a solution to such a problem can be effectively constructed from a 
proof of $F$.
So, for example, once a formula $\ada x\ade y\ p(x,y)$ has been proven, we would know that, for every 
$x$, a $y$ with $p(x,y)$ not only exists, but can be algorithmically found; furthermore, we would be able to actually construct such an algorithm. Similarly, a reduction --- in the sense of Definition \ref{feb2}(3) --- of the 
acceptance problem to the halting problem would automatically come with a proof of 
$\ada x\ada y\bigl(\mbox{\em Halts}(x,y)\add\gneg\mbox{\em Halts}(x,y)\bigr)\mli
\ada x\ada y\bigl(\mbox{\em Accepts}(x,y)\add\gneg\mbox{\em Accepts}(x,y)\bigr)$ in such a theory. 
Does not this look like  exactly what the constructivists have been calling for?..\vspace{-2pt}

\noindent \begin{center} * \ * \ *\vspace{-2pt} \end{center}
 
As a conclusive remark, the author wants to point out that the story told in this paper was only about the tip of the iceberg called CL. Even though the phrase ``{\em the} language of CL" was used in some semiformal contexts, such a language has no official boundaries and, depending on particular needs or taste, remains open to various sorts of interesting new operators.
The general framework of CL is also ready to accommodate any reasonable weakening modifications of its absolute-strength  computation model HPM,\footnote{Among the most natural modifications of this sort might be depriving the HPM of its infinite work tape, leaving in its place just a write-only buffer where the machine constructs its moves. In such a modification the exact type of read access to the run and valuation tapes becomes relevant, and a reasonable restriction would apparently be to allow --- perhaps now multiple --- read heads to move only in one direction. An approach favoring 
this sort of machines would try to model Turing (unlimited) or sub-Turing (limited) computational resources such as memory, time, etc. as games, and then understand computing a problem $A$ with resources represented by $R$ as 
computing  $R\mli A$, thus making explicit not only trans-Turing (incomputable) resources 
as we have been doing in this paper, but also all of the Turing/sub-Turing resources needed or allowed for computing $A$, 
--- the resources that the ordinary HPM, PTM or Turing machine models take for granted. 
So, with $T$ representing the infinite read/write tape as a computational resource, 
computability of $A$ in the old sense would mean nothing but computability of $T\mli A$ in the new sense: 
having $T$ in the antecedent would amount to having infinite memory, only this time provided externally (by the environment)
via the run tape rather than internally via the work tape.}
thus keeping a way open for 
studying logics of sub-Turing computability and developing a systematic theory of interactive complexity.

\end{document}